\documentclass[twocolumn,prd,nofootinbib,showpacs]{revtex4}
\usepackage{graphicx}
\usepackage{color}
\usepackage{amsmath,amssymb,amsfonts,dsfont}
\usepackage{multirow}
\usepackage{ulem}

\pdfoutput=1 

\pdfoutput=1 

\usepackage{lineno}

\newcommand{\lsim}{\mathrel{\mathop{\kern 0pt \rlap
  {\raise.2ex\hbox{$<$}}}
  \lower.9ex\hbox{\kern-.190em $\sim$}}}
\newcommand{\gsim}{\mathrel{\mathop{\kern 0pt \rlap
  {\raise.2ex\hbox{$>$}}}
  \lower.9ex\hbox{\kern-.190em $\sim$}}}

\newcommand{\fermilat}[1]{\textit{Fermi}-LAT {#1}}  
  
\definecolor{grey}{rgb}{0.52, 0.52, 0.51}

\definecolor{purple}{RGB}{128, 0, 128}

\def \aap  {A\&A}

\def \apjs  {ApJS}
\def \apjl  {ApJL}

\begin{document}

\footnotemark

\preprint{TTK-20-46}

\title{Investigating $\gamma$-ray halos around three HAWC bright sources in {\it Fermi}-LAT data}

\author{Mattia Di Mauro,}\email{dimauro.mattia@gmail.com}
\affiliation{Istituto Nazionale di Fisica Nucleare, Sezione di Torino, Via P. Giuria 1, 10125 Torino, Italy}
\affiliation{}
\author{Silvia Manconi}\email{manconi@physik.rwth-aachen.de}
\affiliation{Institute for Theoretical Particle Physics and Cosmology, RWTH Aachen University, Sommerfeldstr.\ 16, 52056 Aachen, Germany}
\author{Michela Negro,}\email{michela.negro@nasa.gov}
\affiliation{NASA Goddard Space Flight Center, Greenbelt, MD 20771, USA}
\affiliation{University of Maryland Baltimore County, Department of Physics, Baltimore, MD 21250, USA}
\author{Fiorenza Donato}\email{donato@to.infn.it}
\affiliation{Dipartimento di Fisica, Universit\`a di Torino, via P. Giuria 1, 10125 Torino, Italy}
\affiliation{Istituto Nazionale di Fisica Nucleare, Sezione di Torino, Via P. Giuria 1, 10125 Torino, Italy}

\begin{abstract}
Numerous extended sources around Galactic pulsars have shown significant $\gamma$-ray emission from GeV to TeV energies, revealing hundreds of TeV energy electrons scattering off of the underlying photon fields through inverse Compton scattering (ICS). 
HAWC TeV gamma-ray observations of few-degree extended emission around the pulsars Geminga and Monogem, and LAT GeV emission around Geminga, suggest that systems older than 100 kyr have multi-TeV $e^\pm$ propagating beyond the SNR-PWN system into the interstellar medium.
Following the discovery of few $\gamma$-ray sources by HAWC at energies E$>100$~TeV, we investigate the presence of an extended $\gamma$-ray emission in \fermilat data around the three brightest sources detected by HAWC up to 100~TeV. We find an extended emission of $\theta_{68} = 1.00^{+0.05}_{-0.07}$ deg around eHWC J1825-134 and $\theta_{68} = 0.71\pm0.10$ deg eHWC J1907+063. The analysis with ICS templates on \fermilat data point to diffusion coefficient values which are significantly lower than the average Galactic one. When studied along with HAWC data, the $\gamma$-ray \fermilat data provide invaluable insight into the very high-energy electron and positron parent populations. 
\end{abstract}

\maketitle

\section{Introduction}
\label{sec:intro}
A new population of very-high-energy (VHE) $\gamma$-ray sources emitting above $56$~TeV has been recently reported by the HAWC observatory \cite{Abeysekara:2019gov}.
All the nine sources are observed as extended in the sky, with angular extension ranging from $0.018$ to $0.52$~deg in radius.
Among them, the sources eHWC J1825-134, eHWC J1907+063 and eHWC J2019+368 continue emitting above 100 TeV, making them the brightest $\gamma$-ray sources along with the Crab nebula at these energies.
The mechanisms producing the observed emission are not yet clear, although a pulsar is found within $0.5$~deg of each source.
These sources could be possible candidates for Galactic cosmic ray (CR) Pevatrons.
The PeV-CRs interacting with the ambient radiation fields are expected to produce hadronic $\gamma$-ray emission, coming from neutral pions which subsequently decay into $\gamma$ rays of energy of about hundreds of TeV \cite{2009ARA&A..47..523H,2013APh....43...71A}.  
$\gamma$ rays at these energies are possibly produced also through leptonic processes, i.e inverse Compton scattering (ICS) of energetic electrons and positrons ($e^\pm$) in the ambient photon fields. 
Both hadronic and leptonic emissions are thought to be produced by CRs in different stages of supernova evolution, namely in their supernova remnants (SNRs), pulsars and pulsar wind nebulae (PWNe) \cite{2018SSRv..214...41B,Gaensler:2006ua,Bykov:2017xpo,2017hsn..book.2159S}. 
The production of $>100$~TeV leptonic emission from ICS in Galactic electron accelerators has been recently reconsidered also in Ref.~\cite{Breuhaus:2020mof}, finding that such emission is possible in the presence of inverse Compton dominated cooling in the source environments.

Numerous extended sources around Galactic pulsars have shown significant $\gamma$-ray emission from GeV to TeV energies, revealing multi-TeV electrons scattering off the underlying photon fields through ICS \cite{Abdalla:2017vci,Abeysekara:2017hyn, Abdollahi_2020}. 
These emissions are typically interpreted as coming from $e^\pm$ confined inside a zone dominated by the influence of the pulsar, thus identified as PWNe, in which the relativistic particle propagation is likely dominated by advection, in particular for young ($t<10$~kyr) objects.
However, when converting the angular extension of the $\gamma$-ray emission to the physical dimensions of the source, this often exceeds the typical scales (few pc) expected for the PWN halo size from hydro-dynamical simulations \cite{Khangulyan:2017pfg,Linden:2017vvb}.
In addition, the $\gamma$-ray emission can be much more extended with respect to the X-ray nebulae corresponding to the same pulsars  \cite{Gaensler:2006ua,2017hsn..book.2159S}. 
The recent observation of few-degree extended $\gamma$-ray emission around nearby pulsars 
Geminga (PSR J0633+1746) and Monogem (PSR B0656+14) at TeV energies in HAWC data \cite{Abeysekara:2017science},  and at GeV energies around Geminga \cite{DiMauro:2019yvh}, has been interpreted as coming from a halo of escaped $e^\pm$, exceeding the PWNe boundaries (TeV halos as named in Ref.~\cite{Linden:2017vvb}, ICS halos in Refs.~\cite{DiMauro:2019yvh,DiMauro:2019hwn}; see also discussion in Ref.~\cite{Giacinti:2019nbu}). 
These observations suggest that the multi-TeV $e^\pm$ producing the $\gamma$-ray emissions for sources older than $10-100$~kyr are probably not confined inside the influence of the SNR-PWN system, but propagate in a region with characteristics similar to the interstellar medium (ISM). In this case, their transport is expected to be dominated by diffusion, rather than advection, as well as by radiative losses. 
This  effect is particularly relevant for evolved objects, such as Geminga and Monogem (342~kyr and 111~kyr, respectively \cite{2005AJ....129.1993M}), for which the pulsar has escaped the parent SNR due to its initial kick velocity \cite{Bykov:2017xpo}. 
Furthermore, these observations indicate that highly energetic $e^\pm$ escaped from their PWNe propagate further in the Galaxy, possibly reaching Earth and contributing to the measured local cosmic-ray fluxes \cite{PhysRevLett.122.041102,Manconi:2020ipm}.
Similar objects have subsequently been identified by HAWC \cite{atel1,atel2}, and many more are expected to be unveiled in present and future $\gamma$-ray observatories \cite{DiMauro:2019hwn,Linden:2017vvb,Sudoh:2019lav}.

Although the transition between the difference evolutionary stages of the SNR-PWN is complex, and the discrimination between TeV/ICS halos and PWNe is still debated, the study of sources with intermediate ages $10-100$~kyr, as the three pulsars considered in this paper, is crucial for  detailed predictions of the expected number of such objects in current and future surveys. 
In Ref.~\cite{Giacinti:2019nbu} the $e^{\pm}$ energy density inside PWNe has been proposed as an estimator for the identification of ICS halos. 
However, the physical extension has been taken from the size of the TeV emission around these objects, which does not necessarily size the ICS halo dimension \cite{DiMauro:2019hwn}.  
By analyzing an extended sample of sources using HESS data we reported evidence that $\gamma$-ray data are well described by an extended halo of $e^\pm$ propagating in a low-diffusion zone around the pulsars, with no evident dependence on the source age \cite{DiMauro:2019hwn}.
The characterization of the extension of these systems along their $\gamma$-ray spectrum is crucial to understand their properties, and to infer the properties of the underlying lepton population, see e.g. \cite{Yuksel:2008rf,Tang:2018wyr,DiMauro:2019yvh,Giacinti:2019nbu,DiMauro:2019hwn}.

In this paper, we search for the {\it Fermi} Large Area Telescope (LAT) ({\it Fermi}-LAT) counterparts of the Galactic $\gamma$-ray sources detected by HAWC at energies E$>100$~TeV \cite{Abeysekara:2019gov}. 
We investigate the presence of extended $\gamma$-ray emission at GeV energies around the three bright sources detected by HAWC up to 100~TeV, for which a detailed spectrum is available.
With respect to standard catalog searches, 
we analyze \fermilat data by using specific physical templates based on the ICS process, which include by construction the energy dependence of the halo extension. 
$\gamma$ rays from GeV to TeV energies are then interpreted in the context of leptonic emission, coming from 
ICS of $e^\pm$ produced and accelerated by PWNe, and propagating in a diffusion-dominated scenario.  
We describe the spectral energy distribution (SED) of these sources from GeV to multi TeV energies, constraining the underlying $e^\pm$ distribution, as well as the transport properties. 

The paper is organized as follows. 
In Sec.~\ref{sec:model} we describe the modeling of the $\gamma$-ray emission of $e^\pm$ from PWNe.
Sec.~\ref{sec:sources} is devoted to a brief description of the three sources in our sample. 
In Sec.~\ref{sec:data} we describe the data selection and the techniques used to explore ICS halos around PWNe in the \fermilat data. 
The results on the optimization of the region of interest around each source are presented in Sec.~\ref{sec:results}. 
Our main results are discussed in Sec.~\ref{sec:ICShalo}, before concluding in Sec.~\ref{sec: conclusions}.

\section{Gamma-rays from electrons and positrons in PWNe}
\label{sec:model} 

We work under the hypothesis that $e^\pm$ pairs accelerated by pulsars and their wind nebulae (PWNe) 
can up-scatter ambient photons to $\gamma$ rays through ICS. We have extensively described the underlying model of this process in Refs.~\cite{DiMauro:2019yvh,DiMauro:2019hwn}, to which we refer for a detailed description. 
Below we describe the main points of our computations.

In the magnetosphere created around Galactic pulsar, $e^\pm$ are produced and likely accelerated at the termination shock, {\it i.e.} where the PWN meets the ISM (see \cite{Gaensler:2006ua,2011ASSP...21..624B,Bykov:2017xpo,Amato:2020zfv} for a careful description of these systems). 
We model the $e^\pm$ spectrum emitted from PWNe $Q(E_e,t)$  by assuming a continuous injection of particles, with a rate following the pulsar spin down energy $L(t)= L_0 /(1+t/\tau_0)^2$ and shaped as \cite{Yuksel:2008rf,Abeysekara:2017science,Tang:2018wyr,DiMauro:2019hwn}:
 \begin{equation}
 Q(E_e, t)= L(t) \left( \frac{E_e}{E_0}\right)^{- \gamma_e} \exp \left(-\frac{E_e}{E_c} \right) \, . 
 \label{eq:Q_E_cont}
\end{equation}
The characteristic pulsar spin-down timescale is set to $\tau_0=12$~kyr following previous papers on a similar topic \cite{Abeysekara:2017science,Tang:2018wyr,DiMauro:2019yvh,Manconi:2020ipm}. We refer to \cite{Manconi:2020ipm} for a comprehensive study on the variation of this parameter and on the effects on the propagated $e^\pm$ flux at Earth.  
The spin-down luminosity $\dot{E}$  of the pulsar is transferred to the $e^\pm$ pairs with an efficiency $\eta$ (see Ref.~\cite{Manconi:2020ipm} for the full set of formulae). The spectral index $\gamma_e$ of high-energy $e^\pm$ can be constrained through observations of PWNe at different wavelengths, in particular in the radio band \cite{Gaensler:2006ua,DiMauro:2019yvh}.

After being produced, $e^\pm$ diffuse in the surrounding medium and lose energy through synchrotron emission in the Galactic magnetic field, and ICS in the interstellar radiation fields (ISRFs). 
Specifically, we solve the transport equation for the  $e^\pm$  number density $\psi = \psi(E, \mathbf{x}, t)\equiv dn/dE$ per unit volume and energy
\begin{equation}
 \frac{\partial \psi}{\partial t}  - \mathbf{\nabla} \cdot \left\lbrace D(E)  \mathbf{\nabla} \psi \right\rbrace + 
 \frac{\partial }{\partial E} \left\lbrace \frac{dE}{dt} \psi \right\rbrace = q(E, \mathbf{x}, t)
 \label{eq:diff}
\end{equation}
following  \cite{2010A&A...524A..51D,DiMauro:2014iia,Manconi:2016byt} (see Ref.~\cite{Manconi:2016byt} for further details). 
Here $D(E)$ is the energy dependent diffusion coefficient, $dE/dt\equiv b(E)$ accounts for the energy losses and $q(E, \mathbf{x}, t)$ is the  $e^-$ and $e^+$ source term. 
The flux of electron $\Phi$ at the Earth is connected to the number density through $\Phi=v /4\pi \;\psi$. 
We include $e^\pm$ energy losses by Inverse Compton  scattering off the ISRF, 
and synchrotron losses on the Galactic magnetic field.
A full-relativistic treatment of Inverse Compton losses has been implemented in the Klein-Nishina regime, according to Ref.~\cite{2010A&A...524A..51D}. 
Since the physical scale of the emission we study (tens of pc) is considerably larger than the pulsar's strong magnetic field region, we consider the magnetic field to be equal to the mean Galactic value $3.6\mu$G \cite{2007A&A...463..993S}. As for the ISRFs, we implement the model in Ref.~\cite{Vernetto:2016alq}, composed by the CMB,  infrared light and starlight. 
The energy losses can be parameterized as $b(E_e)=b_0E_e^2$, with the normalization $b_0$  encoding effectively the synchrotron and ICS intensity losses. 
We consider possible source-by-source variations  of the energy loss properties - given by different magnetic field values  or variations in the ISRF densities - in an effective way, by changing the normalization $b_0$. 

%
The diffusion coefficient in the halo around pulsars is parameterized as $D(E_e)=D_0E_e^{\delta}$, with $D_0=D$(1 GeV) and $\delta=0.33$.  
In light of the recent evidence \cite{Abeysekara:2017science,DiMauro:2019yvh,DiMauro:2019hwn} for a tens of pc extended region around the pulsar where  diffusion is inhibited with respect to the typical values derived for the Galaxy \cite{Kappl:2015bqa,Genolini:2019ewc}, 
we will provide the results as a function of $D_0$.

As already mentioned, the GeV-TeV $\gamma$ rays observed in the halos around pulsars and their PWNe are believed to be produced by the ICS of $e^\pm$ off the ISRF \cite{Vernetto:2016alq}.
The $\gamma$-ray flux produced at a $\gamma$-ray energy $E_{\gamma}$ within a solid angle $\Delta \Omega$ around the source line-of-sight $s$  is computed with a fully numerical approach as:
\begin{eqnarray}
\label{eq:phflux}
 && \Phi_{\gamma} (E_{\gamma}, \Delta \Omega)= \\ \nonumber
&& \frac{1}{4\pi} \int_{m_e c^2}^{\infty} dE_e \int_{\Delta \Omega}  d\Omega \int_0^{\infty} d s \,  
 \mathcal{N}_e (E_e,s,T) \mathcal{P}(E_e, E_{\gamma})\,,
\end{eqnarray}
where  $\mathcal{P}(E_e, E_{\gamma})$ is the power emitted in ICS photons by a single $e^-, e^+$ with energy $E_e$. 
Our implementation for the $\gamma$-ray and  $e^\pm$ flux computations has been extensively validated and compared with other works in  Refs.~\cite{DiMauro:2019yvh,DiMauro:2019hwn}, to which we refer for a detailed discussion of the effect of different assumptions, notably the energy losses and the spectral shape assumed for the  $e^\pm$ pairs.

\begin{figure}[t]
\includegraphics[width=0.5\textwidth]{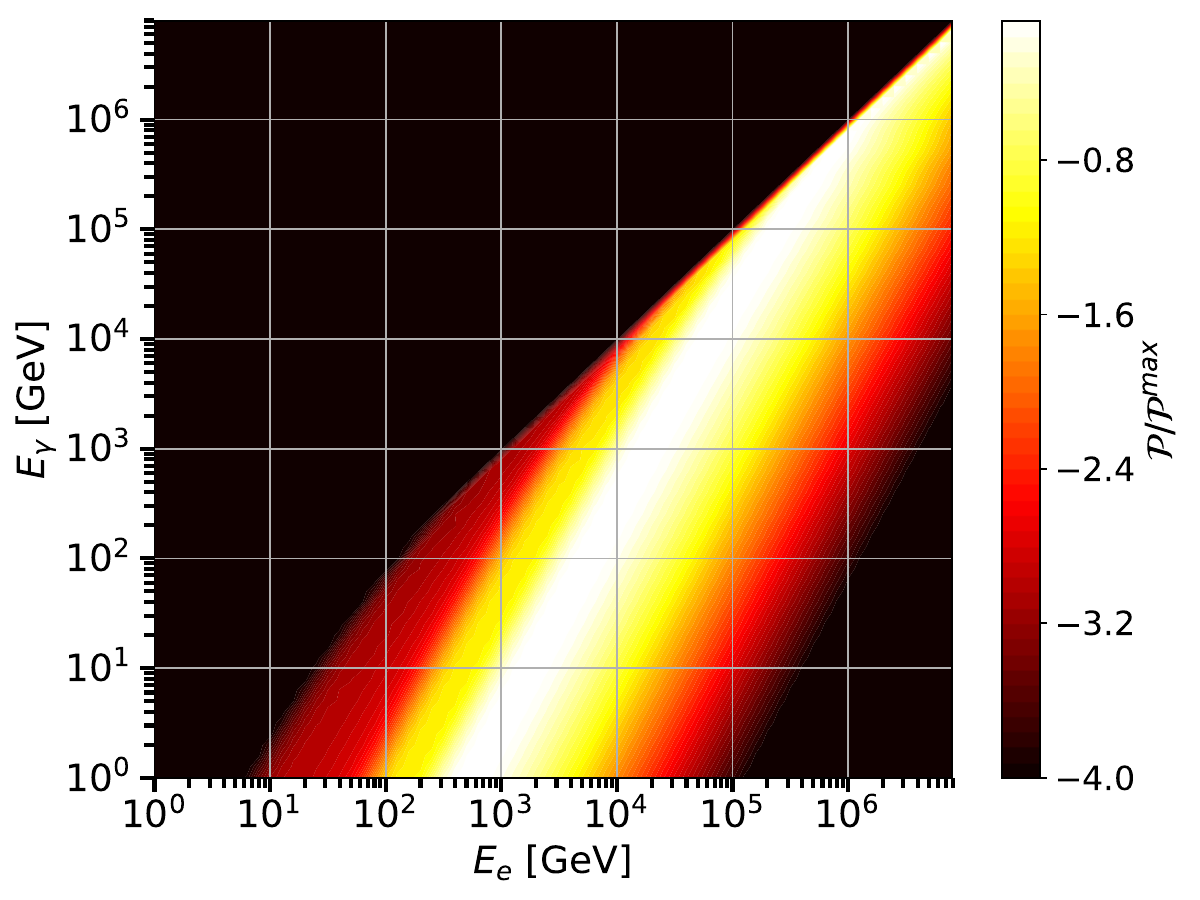}
\caption{Ratio $P/P^{\rm max}$ between the inverse Compton power $P(E_e, E_{\gamma})$ and its maximum value $P^{\rm max}$,  
varying $E_{\gamma}$ - $E_e$. The colors (in logarithmic scale) indicate the $P/P^{\rm max} $values. The maximum of 
$P/P^{\rm max}$ corresponds to white regions in the plot. }  
\label{fig:icpower}
\end{figure}
In Fig.~\ref{fig:icpower} we display the ratio $P/P^{\rm max}$ between the inverse Compton power $P(E_e, E_{\gamma})$ and its maximum value $P^{\rm max}$, varying $E_{\gamma}$ - $E_e$ values and for the ISFR as in \cite{Vernetto:2016alq}. 
This figure is meant to show the $e^\pm$ energies which more likely correspond to an ICS $\gamma$-ray photon. 
The logarithmic color scale indicates the value of  $P(E_e, E_{\gamma})$/$P^{\rm max}$. White regions correspond to 
$P(E_e, E_{\gamma})/P^{\rm max} \simeq 1$, namely to the bulk of  $E_e$ originated by a $\gamma$ ray with $E_{\gamma}$.
A $\gamma$ ray with $E_{\gamma} = 0.1/1/10/100$ TeV is mainly produced by  $e^\pm$ with $E_e\sim 6/20/70/300$ TeV.
From this plot, we can also read the typical $E_e$ which corresponds to the $\gamma$-ray emission seen by HAWC at $>56-100$~TeV. 
For $E_{\gamma} \sim 10$~TeV, the peak of the ICS emission corresponds to $E_e=30-150$~TeV. 
An important contribution to the photons observed by HAWC  comes from $e^\pm$ with energies of hundreds up to thousands of TeV. Therefore, HAWC observations of such very-high energy photons could be a probe of the fact that PWNe are pevatron accelerators.
As for the \fermilat energy range $E_\gamma=1-100$~GeV, this telescope probes the population of ICS $e^\pm$ with a peak energy between 1 and 10 TeV.

As extensively discussed in Refs.~\cite{DiMauro:2019yvh,DiMauro:2019hwn,Zhang:2020vga}, the pulsar proper motion can significantly shape the morphology of the observed ICS emission at GeV energies. We verified that the effect on the observed surface brightness for our set of sources is negligile, and smaller than the typical uncertainties in the measured source extension, see Appendix~\ref{app:pmotion}.

\section{HAWC bright sources at $E> 100$ TeV}
\label{sec:sources} 

Among the sources detected by HAWC at energies larger than $56$~TeV in Ref.~\cite{Abeysekara:2019gov}, we select the sources eHWC J1825-134, eHWC J1907+063 and eHWC J2019+368, which exhibit the most significant emission at $E_{\gamma}>100$~TeV. These are also the only three sources in Ref.~\cite{Abeysekara:2019gov} for which the SED between 1 and $\sim$100 TeV has been published.
An energetic pulsar from the ATNF catalog \cite{2005AJ....129.1993M} is found within angular distances about $0.3$~deg from the peak of the HAWC emission in all three cases. 
The main characteristics of the three pulsars are reported in Tab.~\ref{tab:sample}, together with the name of the HAWC sources and the angular extension detected at $>56$~TeV calculated as the $68\%$ containment radius ($\theta_{68}$) \cite{Abeysekara:2019gov}. 
The HAWC Collaboration released also the angular extension in the entire energy range analyzed. These are $0.53 \pm0.02$,  $0.67 \pm0.03$ and $0.30 \pm0.02$, respectively. These angular sizes, as expected are slightly larger than the one given for $E>56$ TeV. 
Since the lower bound of the entire energy range used to measure the extension is not clearly stated, we will use in the paper the values measured for E > 56 TeV.
In the last column we report an estimate for the $\theta_{68}$ at $E_{\gamma}=10$~GeV.
We will always report in this paper the size of extension as $\theta_{68}$.
We down-scale the source extensions at GeV energies from the extensions observed by HAWC at TeV, following the ICS model previously described, which predicts the evolution of the ICS halo extension as a function of the energy. We fix $\gamma_e=1.8$, while $D_0$ is derived in order to match the HAWC observations at energies $>56$~TeV. This value of $\gamma_e$ is representative of source SED and changing it to slightly different numbers is not going to change the extension.
The predictions for the size of extension at 10 GeV are between 0.20-0.65$^\circ$ making these sources suitable for the search also in LAT data.

\begin{table*}
\begin{center}
\begin{tabular}{|c|c|c|c|c|c|c|c|c|c|}
\hline
PSR name  &   $l$  &  $b$  &   $d$ & $T$ & $\dot{E}$ &  HAWC source  &  $\theta_{68}$ at $E>56$~TeV & $\theta_{68}$ at $E=10$~GeV \\ 
\hline
  &	[deg]	 &  [deg]&   [kpc] &  [kyr]        &   [erg/s]   &   & [deg]  &  [deg]    \\ 
\hline
  J1826-1334 &  $18.00$ & $0.69$ &  $3.61$ & $21.4$ &  $ 2.8\cdot 10^{36}$  & eHWC J1825-134 & $0.36\pm 0.05$ &   $0.50$ \\
 J1907+0602  &  $40.18$ & $-0.89$ &  $2.37$ & $19.5$ &  $ 2.8 \cdot 10^{36}$  & eHWC J1907+063 & $0.52 \pm 0.09$ &   $0.65$ \\
 J2021+3651  &  $75.22$ & $0.11$ &  $1.8$ & $17$ &  $ 3.4\cdot 10^{36}$  & eHWC J2019+368 & $0.20\pm0.05$ &   $0.23$ \\
\hline
\end{tabular}
\caption{Very-high energy $\gamma$-ray sources detected by HAWC and analyzed in this work with \fermilat data. 
The columns contain for each source: the name of the pulsar found within small angular distances to the HAWC source, its Galactic longitude ($l$), latitude ($b$), distance ($d$) and spin-down age ($T$) as found in the ATNF catalog; the name of the associated HAWC source in Ref.~\cite{Abeysekara:2019gov} (eHWC for high-energy threshold HAWC), along with the measured angular extension at energies $>56$ TeV given as $\theta_{68}$ and the estimated $\theta_{68}$ for one representative value in the {\it Fermi}-LAT energy range ($10$~GeV, see text for more details).}
\label{tab:sample}
\end{center}
\end{table*}

In Appendix \ref{sec:sourceJ1825}-\ref{sec:sourceJ2019}  we describe the three sources and briefly review their multi-wavelength observations.

\section{Data selection and analysis techniques}
\label{sec:data} 
In this section we describe the data selection and the techniques exploited to  study the extended ICS halos around PWNe in the \fermilat data.

We analyze eleven years\footnote{Mission Elapsed Time (MET): 239557417 s $-$ 586490000 s} of \fermilat data exploiting the latest release of Pass 8 data processing (P8R3) by means of the publicly available \textit{fermitools} \cite{2018arXiv181011394B}. We select SOURCEVETO\footnote{This new event class maximizes the acceptance while minimizing the irreducible cosmic-ray background contamination. To compare with previous data releases, SOURCEVETO class has the same contamination level of P8R2\_ULTRACLEANVETO\_V6 class while maintaining the acceptance of P8R2\_CLEAN\_V6 class.} class events (FRONT+BACK type), passing the basic quality filter cuts\footnote{DATA\_QUAL$>$0 \&\& LAT\_CONFIG==1}. 
The energy dispersion is taken into account through the dedicated \textit{Fermipy} routine, and the P8R3\_SOURCEVETO\_V2 response functions are used to analyze the data. 
A standard cut selecting zenith angles $<105^\circ$ is applied in order to exclude the Earth Limb's contamination. The nominal energy range of our analysis selects events with reconstructed energy between 1 GeV and 1 TeV (except for eHWC J2019+368, for which we select $\gamma$-rays above 6 GeV, see Sec. \ref{sec:results}).

For each source we use the public \textit{Fermipy} package (version 0.18.0) to perform a binned analysis with eight bins per energy decade. We analyze the $14\times14$ deg$^2$ regions of interest (ROI) centered in the source positions reported by HAWC (see Table \ref{tab:sample}) and choose pixel size of $0.08$ deg. We make an exception for eHWCJ1825-137, opting for repeating the analysis twice: 1) centering the ROI in the position of the associated pulsar, and 2) centering the ROI in the position of the $\gamma$-ray emission peak, as reported in the 4FGL catalog \cite{Abdollahi_2020}. 

The general procedure adopted for the three sources can be summarized in three main steps: (i) realization of the ICS templates for different $D_0$ values, (ii) baseline analysis devoted to the ROI optimization, (iii) iterative procedure to scan the $D_0$ parameter space and build the associated likelihood profile. Steps (ii) an (iii) are repeated for 3 different interstellar emission models (IEMs): (1) the latest released \textit{official} IEM, namely {\tt gll\_iem\_v07.fits}\footnote{A complete discussion about this new IEM can be found at \url{https://fermi.gsfc.nasa.gov/ssc/data/analysis/software/aux/4fgl/Galactic_Diffuse_Emission_Model_for_the_4FGL_Catalog_Analysis.pdf}}, used in the the 4FGL catalog production, and hereafted referred as IEM-\textit{4FGL}; (2) the IEM employed in the analysis of the \textit{Galactic center} excess \cite{TheFermi-LAT:2017vmf} and hereafter addressed as IEM-\textit{GC}; (3) an \textit{alternative IEM} (hereafter labelled as IEM-\textit{ALT1}) used (along with eight other models produced varying CR propagation properties) in the first \fermilat SNR catalog \cite{Acero:2015prw} to explore the systematic effects associated with the choice of the IEM. 
We show results with only one of the 8 IEM models created in \cite{Acero:2015prw}, since we tested that by running the analysis for the others we find very similar results with respect to IEM-\textit{ALT1}.
As for the isotropic emission contribution, the models employed in this study are {\tt iso\_P8R3\_SOURCEVETO\_V2\_v1.txt} when the analysis involves the IEM-\textit{4FGL}, and other two different isotropic emission models associated with IEM-\textit{GC} and IEM-\textit{ALT1}.
Here we  detail the three main steps of the source analysis. They are common to all three sources, even if some aspects (e.g. the energy range and the center of the ROI) will be optimized in each cases (see Sec. \ref{sec:results} for the details).

\begin{description}
\item[(i) Creation of ICS templates $-$] We generate the ICS templates following the same procedure used in \cite{DiMauro:2019yvh} and briefly outlined in Sec~\ref{sec:model}. The extension of the ICS halo depends mostly on the diffusion coefficient $D_0$: the larger the value of $D_0$, the more extended is the ICS halo. For each HAWC source we produce 30 ICS templates by varying $D_0$ from $10^{25}$ to  $10^{28}$  cm$^2$/s (in logarithmic spacing).

\item[(ii) ROI model optimization $-$]  The generic model for each ROI consists of the interstellar emission, the isotropic emission, the ICS template, and the list of sources that populate the ROI according to the 4FGL catalog. In particular, we include and leave free in the fit all the sources in a square $18\times18$ deg$^2$ centered in the ROIs. The optimization runs in a multi-parameter fitting procedure in which the free parameters are the source SED parameters (according to the parametrization of the 4FGL catalog), the normalization and the spectral index of the IEM, the normalizations of the isotropic emission and of the ICS templates.
The best-ft values of the SED parameters are comparabile with the initial ones taken from the 4FGL.

To better study the properties of the ICS halo we perform a double step study. First we test the presence of an extended emission using the geometrical models provided by the Fermitools, namely a \textit{uniform disk} and a \textit{2D Gaussian} template. This is done by using the {\tt gta.extension} tool implemented in {\tt Fermipy} that performs a re-localization of the source and search for a spatial extension at the same time. Secondly, we substitute the geometrical model with the ICS template obtained as described in step (i). If the ICS process is indeed the primary mechanisms responsible for $\gamma$-ray extended emission from our sources, the ICS templates should provide a better fit to the data than the geometrical ones.

The multi-parameter fitting procedure proceeds by few steps. We perform a first fit to the data with all the parameters free to vary using the {\tt gta.fit}. Then, we remove from the model sources detected with a test statistic ($TS$)\footnote{TS is defined as $-2ln(\mathcal{L}_0/\mathcal{L})$, $\mathcal{L}_0$ being the likelihood of the null hypothesis (no source is present) and $\mathcal{L}$ the likelihood when including the source in the model. If the Wilks' theorem \cite{Wilks1938} applies, as in our case, a TS=25 corresponds to $\sim5\sigma$ detection significance.} lower than 25. We perform a second fit with the remaining components, and then we search for new sources with a TS$>$25 within 5 degrees from the center of the ROI (using the {\tt Fermipy} {\tt gta.find\_source} routine). We include the new sources in the model, and perform a final fit with all the SED parameters of the sources, the IEM and isotropic template free to vary. None of the new sources are within $1^{\circ}$ of the sources of interest and their presence do not affect our results. For each HAWC source, this baseline procedure is repeated for each of the three Galactic IEMs described above. 
As detailed in the following Section, this step includes slightly different procedures for each ROI.

\item[(iii) Scan in $D_0$ $-$] 
Once the model for the considered ROI has been optimized, it is then exploited to produce a set of 30 new models differing only for the ICS template (obtained, as described in step 1, varying the diffusion parameter). A further fit is then performed in which the normalizations of the ICS, the IEM and the isotropic emission templates are left free to vary. The likelihood value obtained for every model is used to derive the likelihood profile as a function of the $D_0$. This scan is repeated for each HAWC source and for all the three galactic IEMs. 
\end{description}

\section{Results on extended $\gamma$-ray emission around the sources of interest}
\label{sec:results}
In this section we present the results of the ROI optimization process to each of our three sources in the search for an extended 
$\gamma$-ray emission. 

\subsection{eHWC J1825-134}
\label{sec:1825opt}
\begin{figure}[t]
\includegraphics[width=0.49\textwidth]{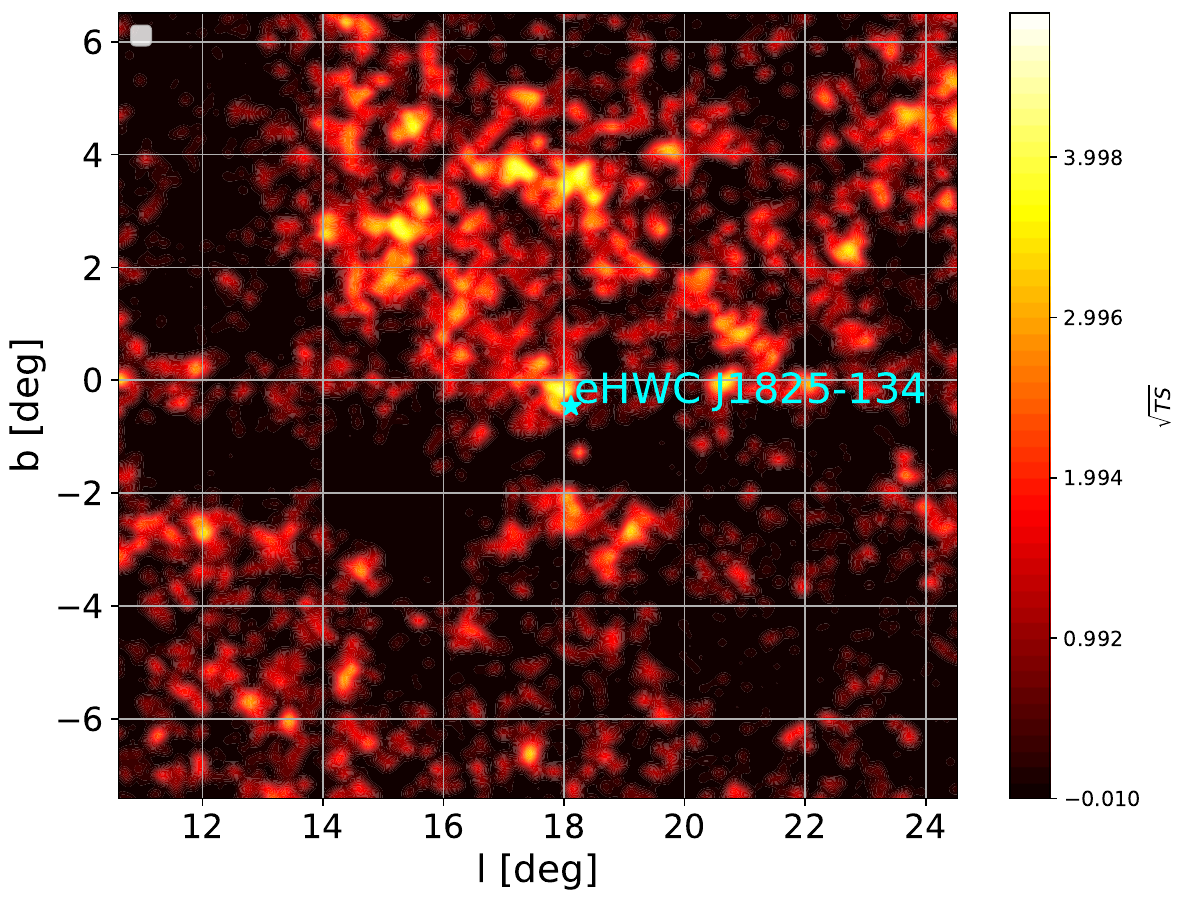}
\includegraphics[width=0.49\textwidth]{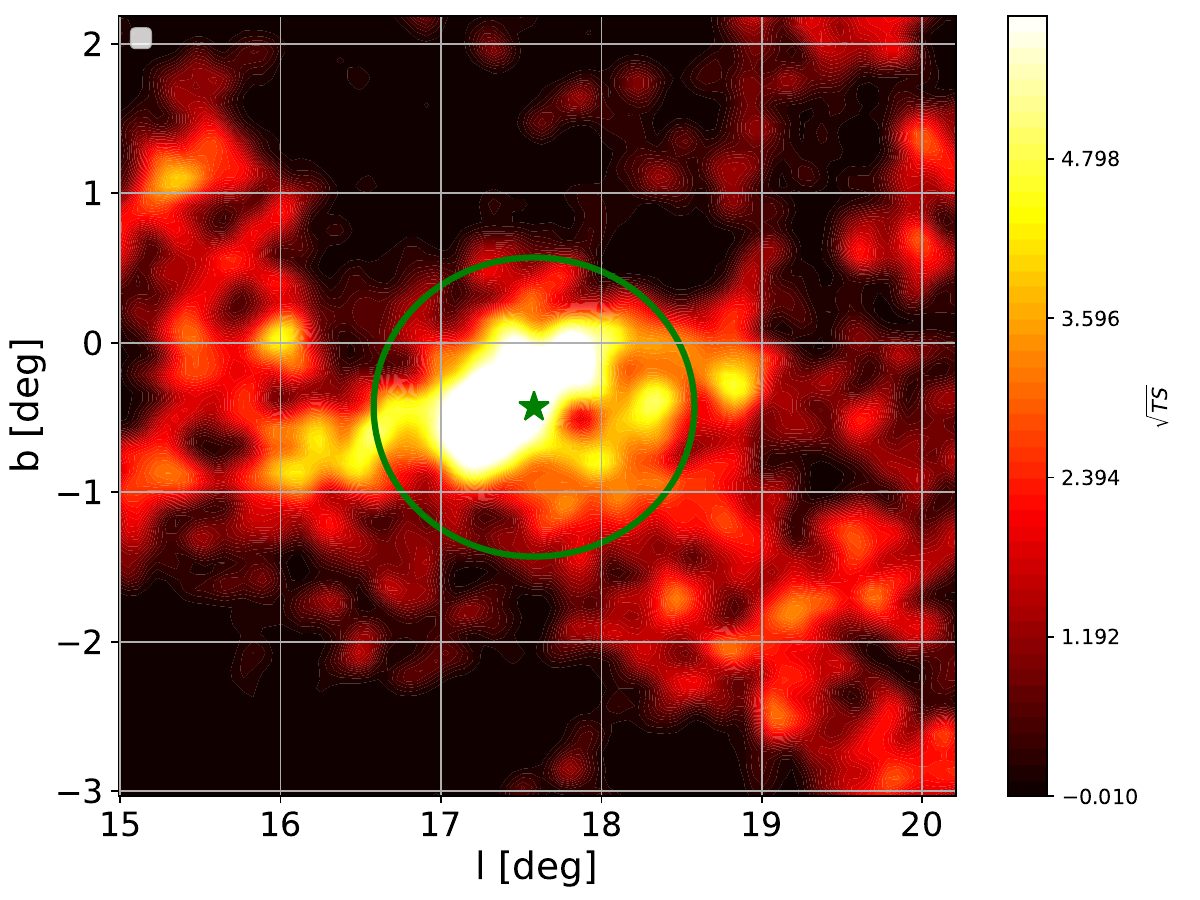}
\caption{Top panel: $\sqrt{TS}$ map of the ROI analyzed in \fermilat data for the source eHWC J1825-134 after subtracting the best-fit model obtained as described in Sec. \ref{sec:data}. The cyan star indicates the position of the eHWC source, while the associated pulsars are located at the ROI center Bottom panel: same as the left panel but without accounting for the source of interest in the model. The green star and circle indicate the best fit position and angular extension of the source as found in our analysis between 1-1000 GeV with a Gaussian template. The maps are smoothed with a Gaussian kernel of $1\sigma$.}
\label{fig:tsmaps_final1825}
\end{figure}

A search for an extended emission for the source eHWC J1825-134 reveals that the Gaussian template is highly preferred over a disk shape ($2\Delta \rm{Log}(\mathcal{L})=170$). Therefore, the results reported in this section are computed with the Gaussian morphology only. Instead, in Sec~\ref{sec:ICShalo} we will use the ICS templates.
When we use the IEM-\textit{4FGL}, the best-fit value for the center of the extended emission is at $l=17.58 \pm 0.03$ deg and $b= -0.43 \pm 0.04$ deg, while the spatial extension given as the $68\%$ containment radius is $\theta_{68} = 1.00^{+0.05}_{-0.07}$ deg. 
In the 4FGL the extended source associated with eHWC J1825-134 is named as 4FGL J1824.5-1351e (see Sec.~\ref{sec:sourceJ1825}), with best fit coordinates for the center of its extension $(l,b)=(17.57,-0.45)$ deg and $1.12$ deg\footnote{The extension is provided in {\tt Fermipy} as the $68\%$ containment radius, which is equivalent to the standard deviation of a 1D Gaussian. Instead, in the 4FGL the source 4FGL J1824.5-1351e is modeled with a 2D Gaussian with standard deviation of $0.75$ deg. Since the standard deviation of a 2D Gaussian is about 1.51 smaller than the correspondent standard deviation of a 1D Gaussian, the extension we find once converted to a 2D Gaussian becomes $0.68^{+0.03}_{-0.05}$ deg, which is compatible with the value reported in the 4FGL.}.
Our results for the position and size of extension are thus perfectly compatible with the results reported in the 4FGL.
The $TS$ of the source, when it is modeled as extended, is 846, so much larger than the result reported in the 4FGL catalog\footnote{In the 4FGL catalog this source is detected with at 19.75$\sigma$ significance. Considering 5 degrees of freedom (2 for the position, two for the SED parameters and the size of extension) this detection significance corresponds to a $TS$ of about 410.}.
Instead, the value found in Ref.~\cite{Principe:2020zqe} is 1040.
We show in Fig.~\ref{fig:tsmaps_final1825} the $TS$ map computed in the ROI after the optimization procedure. 
This map is produced by calculating the $TS$ of an additional source with a power-law spectrum and index 2.0 located at the different pixels of the ROI. There are not particularly large residuals and the highest peaks in $\sqrt{TS}$ are about $4\sigma$.

This source is thought to be powered by the pulsar J1826-1334 (see Sec.\ref{sec:sourceJ1825}). Its emission is very bright both at {\it Fermi}-LAT and VHE energies, and its extension has been found to strongly evolve with energy both in HESS data \cite{Aharonian:2006zb,Abdalla:2018qgt} ($0.8$ deg at 500 GeV and $0.2$ deg at 30 TeV ) and in {\it Fermi} LAT data \cite{Principe:2020zqe} (from $1.5$ deg at GeV energies to $1.0$ deg at hundreds of GeV).

\begin{table*}
\begin{center}
\begin{tabular}{|c|c|c|c|c|c|c|}
\hline
 IEM-\textit{4FGL}  &   $1-3$ GeV  &  $3-10$ GeV  &  $10-30$ GeV & $30-100$ GeV & $100-1000$ GeV \\ 
\hline
\hline
 $l$ [deg] & $17.43\pm0.08$  & $17.61\pm0.08$  &  $17.45\pm0.09$ &
  $17.56\pm0.07$ &  $17.57\pm0.07$ \\
   $b$ [deg] & $0.25\pm0.08$  & $-0.64\pm0.10$  &  $-0.61\pm0.10$ &
  $-0.69\pm0.08$ &  $-0.66\pm0.05$ \\
  $\theta_{68}$ [deg] &  $0.93\pm0.13$ & $1.27\pm0.15$ & $1.46\pm0.08$  & $1.07\pm0.11$ & $0.64\pm0.06$ \\  
   $TS$($TS_{\rm{ext}}$)  & 251(32)  & 327(150)  &  350(213) &  278(178) &  179(110) \\
\hline
 IEM-\textit{GC}  &   $1-3$ GeV  &  $3-10$ GeV  &  $10-30$ GeV & $30-100$ GeV & $100-1000$ GeV \\ 
\hline
\hline
 $l$ [deg] & $17.45\pm0.07$  & $17.49\pm0.06$  &  $17.50\pm0.07$ &
  $17.57\pm0.08$ &  $17.57\pm0.06$ \\
   $b$ [deg] & $0.16\pm0.06$  & $-0.12\pm0.05$  &  $-0.36\pm0.08$ &
  $-0.64\pm0.08$ &  $-0.63\pm0.07$ \\
   $\theta_{68}$ [deg] &  $0.79\pm0.10$ & $1.08\pm0.08$ & $1.13\pm0.13$  & $1.05\pm0.06$ & $0.69\pm0.12$ \\  
 $TS$($TS_{\rm{ext}}$)  & 743(52)  & 564(222)  &  300(175) &  255(158) &  190(115) \\
 \hline
  IEM-\textit{ALT1}  &   $1-3$ GeV  &  $3-10$ GeV  &  $10-30$ GeV & $30-100$ GeV & $100-1000$ GeV \\ 
  \hline
 $l$ [deg] & $17.62\pm0.09$  & $17.64\pm0.07$  &  $17.60\pm0.07$ &  $17.62\pm0.07$ &  $17.57\pm0.06$ \\
 $b$ [deg] & $0.08\pm0.06$  & $-0.23\pm0.07$  &  $-0.50\pm0.09$ &  $-0.73\pm0.07$ &  $-0.67\pm0.06$ \\ 
 $\theta_{68}$[deg] &  $0.75\pm0.11$ & $1.06\pm0.08$ & $1.23\pm0.13$  & $0.99\pm0.08$ & $0.66\pm0.04$ \\  
 $TS$($TS_{\rm{ext}}$)  & 704(42)  & 369(165)  &  279(164) &  220(149) &  193(120) \\
\hline
\hline
\hline
\end{tabular}
\caption{Best-fit values for the position, extension and significance of detection for eHWC J1825-134, as found analyzing {\it Fermi}-LAT data in different energy bins from 1 to 1000 GeV. These results are expressed with the longitude ($l$) and latitude ($b$), $68\%$ containment angle $\theta_{68}$, the $TS$ for the detection of the source and $TS$ of extension ($TS_{\rm{ext}}$). Each row block corresponds to  the three IEM considered in this paper.}
\label{tab:1826}
\end{center}
\end{table*}

The extension of these sources is expected to be between $0.3^{\circ}-0.6^{\circ}$ (see Tab.~\ref{tab:sample}). These sizes are much smaller than the PSF size at about 1 GeV, which is about $1^{\circ}$. However, $0.3^{\circ}-0.6^{\circ}$ is larger than the typical precision at which the position of a source in the 4FGL is detected that is between $0.05^{\circ}$ for the brightest and $0.2^{\circ}$ for the faintest sources.
As already mentioned, in this work we make use of a physically-motivated ICS template which intrinsically includes the energy dependence of the halo extension around a PWN. 
However, we check the geometrical Gaussian template in the \fermilat data, as in Ref.~\cite{Principe:2020zqe}.

We run the analysis of extension and localization in the following energy bins: 1-3, 3-10, 10-30, 30-100 and 100-1000 GeV 
 using the IEM-\textit{4FGL}, IEM-\textit{ALT1} and IEM-\textit{GC}.
 In our results (Tab.~\ref{tab:1826}) the extension shows an evolution with energy similar to Refs.~\cite{Principe:2020zqe, Abdalla:2018qgt}.
 Since a different definition of the size of extension is used by HESS \cite{Abdalla:2018qgt}, we apply the correction factor in Eq.~4 of Ref.~\cite{Principe:2020zqe}\footnote{The HESS Collaboration published the extension as the radial distance at which the emission in the southern half of the nebula drops to a factor $1/e$ relative to the maximum, starting from the position of the pulsar PSR J1826-1334 \cite{Abdalla:2018qgt}. We do not correct for the different position we are assuming with respect to Ref.~\cite{Abdalla:2018qgt} but the difference above 100 GeV is just about $0.1-0.2^{\circ}$. Moreover, the statistical errors of HESS data, that are of the order of $20\%$ ($>30\%$) below (above) 1 TeV, are larger than this effect.}.
However we do not correct for the different central position and the projected extension along the preferred source emission direction as done in the H.E.S.S paper and in \cite{Principe:2020zqe}.
 The value of the weighted average of the extension found among the different IEMs, and calculated by using the error as the weight is: $0.81\pm0.07,1.09\pm0.05,1.33\pm0.06, 1.04\pm0.04, 0.66\pm0.03$ deg ( $0.81\pm0.12,1.09\pm0.18,1.33\pm0.20, 1.04\pm0.05, 0.66\pm0.03$ deg if we use the difference between the average values as systematics uncertainties). 
The extension thus increases from 1-3 GeV to 3-10 GeV, compatible with \cite{Principe:2020zqe} (considering the systematics on the value of $\theta_{68}$ due to the choice of the IEM model), and decreases at higher energies, being compatible with the measurement reported with HESS data at energies $>100$ GeV \cite{Abdalla:2018qgt}.

\begin{figure*}[t]
\includegraphics[width=0.49\textwidth]{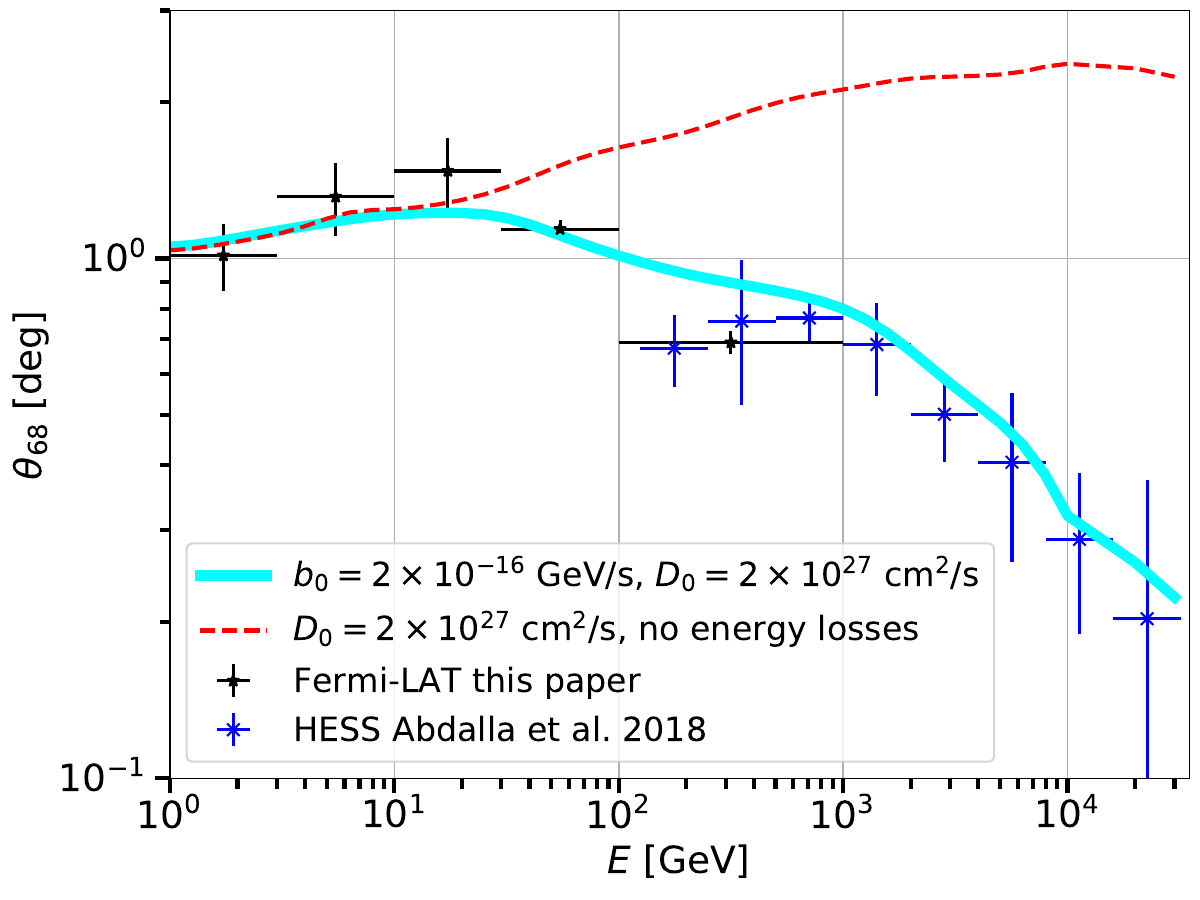}
\includegraphics[width=0.40\textwidth]{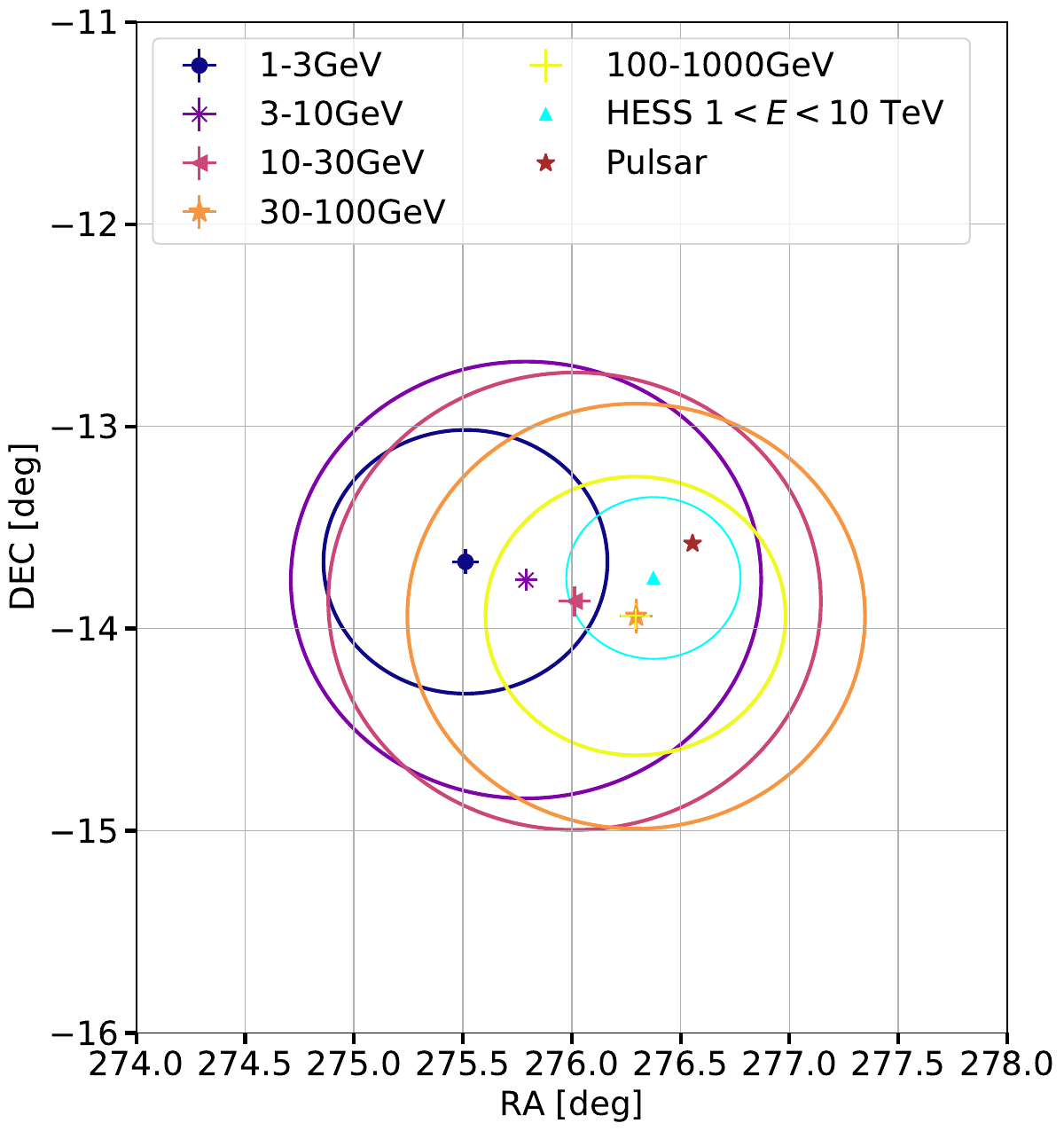}
\caption{Upper Panel: Weighted average for the extension of the $\gamma$-ray flux around the source eHWC J1825-134 (black data) as measured in our analysis of {\it Fermi}-LAT data using a gaussian template with three IEM models (see values in Tab.~\ref{tab:1826}), shown along the analysis of HESS data \cite{Abdalla:2018qgt}. 
We also report the ICS predictions (cyan solid line) fixing $D_0=2\times 10^{27}$ cm$^2$/s and $b_0 =2\times 10^{-16}$ GeV/s (see Fig.~\ref{fig:likelihood}) and the case where only diffusion is considered in the calculation (red dashed line).
Lower Panel: Best-fit position obtained in different energy bins on \fermilat data. The center (size) of the circles describes the best-fit ($1\sigma$ error) of the position. We also display the best-fit position reported in \cite{Abdalla:2018qgt} between $1-10$ TeV, and the position of the pulsar as found in the ATNF catalog.}  
\label{fig:ext_final1825}
\end{figure*}

We test if the peculiar trend of  $\theta_{68}$ as a function of energy is compatible with $e^{\pm}$ injected by the PWN and losing energy for synchrotron radiation and ICS, while diffusing in a low-diffusion bubble located around the source.
We parameterized energy losses as $2\times 10^{-16} E^2$ GeV/s and the diffusion coefficient of  $2\times 10^{27}$ cm$^2$/s, which are the best-ft values we will find by fitting the flux as a function of energy as resulting from \fermilat and HESS data (see Sec.~\ref{sec:ICShalo}).
The best-fit position we find with a Gaussian function evolves with energy (see Tab.~\ref{tab:1826} and the right panel of Fig.~\ref{fig:ext_final1825}). Instead, the ICS model position is energy independent. Therefore, we decide to rerun the extension analysis fixing the position of the Gaussian template to the best-fit we find above 30 GeV. The best-ft for $\theta_{68}$ changes by $20-25\%$ below 10 GeV and $10\%$ between 10-30 GeV. This is expected since the offset with respect to the best-fit position at $E>30$ GeV is larger at lower energies (see right panel of Fig.~\ref{fig:ext_final1825}). We locate the position of the ICS template at the same position of the Gaussian function. This choice makes the comparison of the extension consistent between the two models.
The values of $\theta_{68}$ as a function of energy are reported in the left panel Fig.~\ref{fig:ext_final1825}. 
They have been obtained from the analysis with a Gaussian template and compared with the prediction from  the ICS model. 
We find that the evolution of $\theta_{\rm 68}$ obtained within the geometrical model, both with \fermilat and HESS data, is compatible with the predictions of the ICS model in the whole energy range, which covers more than 4 decades in energy.
We also show the extension obtained as a function of energy when we include only the diffusion process in the calculation (see Eq.~\ref{eq:diff}). In this case the value of $\theta_{\rm 68}$ steadily increases with energy and reaches a plateau above 1 TeV. The contribution from diffusion explains the observations for the extension below 10 GeV but above these energies the addition of the energy losses is needed to follow the decreases shape of the $\theta_{\rm 68}$ data. The addition of the energy losses decreases the extension at energy larger than 10 GeV, since in this regime losses become more important than diffusion. This causes the $e^{\pm}$ in the surrounding of the source to travel shorter distances before losing most of their energies.

From the right panel of Fig.~\ref{fig:ext_final1825}, we note that the best-fit for the center of the extended emission shows an evolution with energy as well. The positions in the two highest energy bins almost coincide, and are compatible with the one reported in \cite{Abdalla:2018qgt} between $1-10$~TeV. The lower energy bins are instead offset with respect to results in the $30-100$ and $100-100$ GeV bins. 
In particular, the best fit position in the $1-3, 3-10, 10-30$ GeV bins is displaced by  $\sim 0.8/0.5/0.3$ deg  with respect to  $E>30$ GeV results.
The evolution of the position as a function of energy in {\it Fermi}-LAT data has been recently reported also in \cite{Principe:2020zqe}, whose results are compatible with what we find in this paper.
The different position with the energy  is hardly explained by the pulsar proper motion, which has a transverse velocity of $440$ km/s, roughly in the direction of the position displacement \cite{Pavlov:2007av}. 
However, the extent of the displacement between 1 GeV and 1000 GeV is roughly $0.8$ deg, as shown in right panel of Fig.~\ref{fig:ext_final1825}. 
For a source like the PSR J1826-1334, located at 3.61 kpc, this would imply a distance traveled in the transverse direction of about 50 pc, that for the age of $T=21.4$ kyr would correspond to $v_T=$ 2300 km/s. 
This value is a factor of 5 larger than the value measured in \cite{Pavlov:2007av}.
In other words, in 21.4 kyrs the angular displacement for a pulsar moving with 440 km/s should be only $0.15$ deg.
In addition, the pulsar proper motion would not explain the upturn to higher declinations of the position in the HESS data and ATNF pulsar. 
The evolution of the position with energy could be due to the interaction of the supernova shock wave with the PWN.
The supernova shock wave could have interacted in one particular direction of the ISM and this could have created a reverse shock that swept out the PWN in the direction of the displacement of the position with energy \cite{Gaensler_2003}.
Since the morphology of eHWC J1825-134 is highly energy dependent, we will  make two different choices for the center of the ICS template. 
The standard approach is to center the template at the position of the $\gamma$-ray peak. 
We also perform the analysis centering, instead, the template at the location of the pulsar, finding similar results.

We do not perform an off-pulsed analysis of this source since the pulsar associated with this source has not been detected by {\it Fermi}-LAT.
However, the pulsar PSR J1826-1256, that is only about $1^{\circ}$ away from eHWC J1825-134, is detected in our analysis with $TS\sim20000$. We perform an off-pulse analysis for this source and verify that we find similar results for the localization and extension of the source eHWC J1825-134. Therefore, the pulsar PSR J1826-1256 does not affect the results for this source.

\subsection{eHWC J1907+063}
The pulsar associated with eHWC J1907+063 (J1907+060) is a very bright source, detected at about 110$\sigma$ significance ($TS=14400$) in the 4FGL.
However, no extended emission has been detected so far around this source, see Appendix \ref{sec:J1907}. 
In Fig.~\ref{fig:tsmaps_final1908} we show the $TS$ map of the ROI after running the optimization explained in Sec.~\ref{sec:data}.
There are no significant residuals in the ROI, meaning that our background model is appropriate for explaining LAT data in this ROI. 
\begin{figure}[t]
\includegraphics[width=0.49\textwidth]{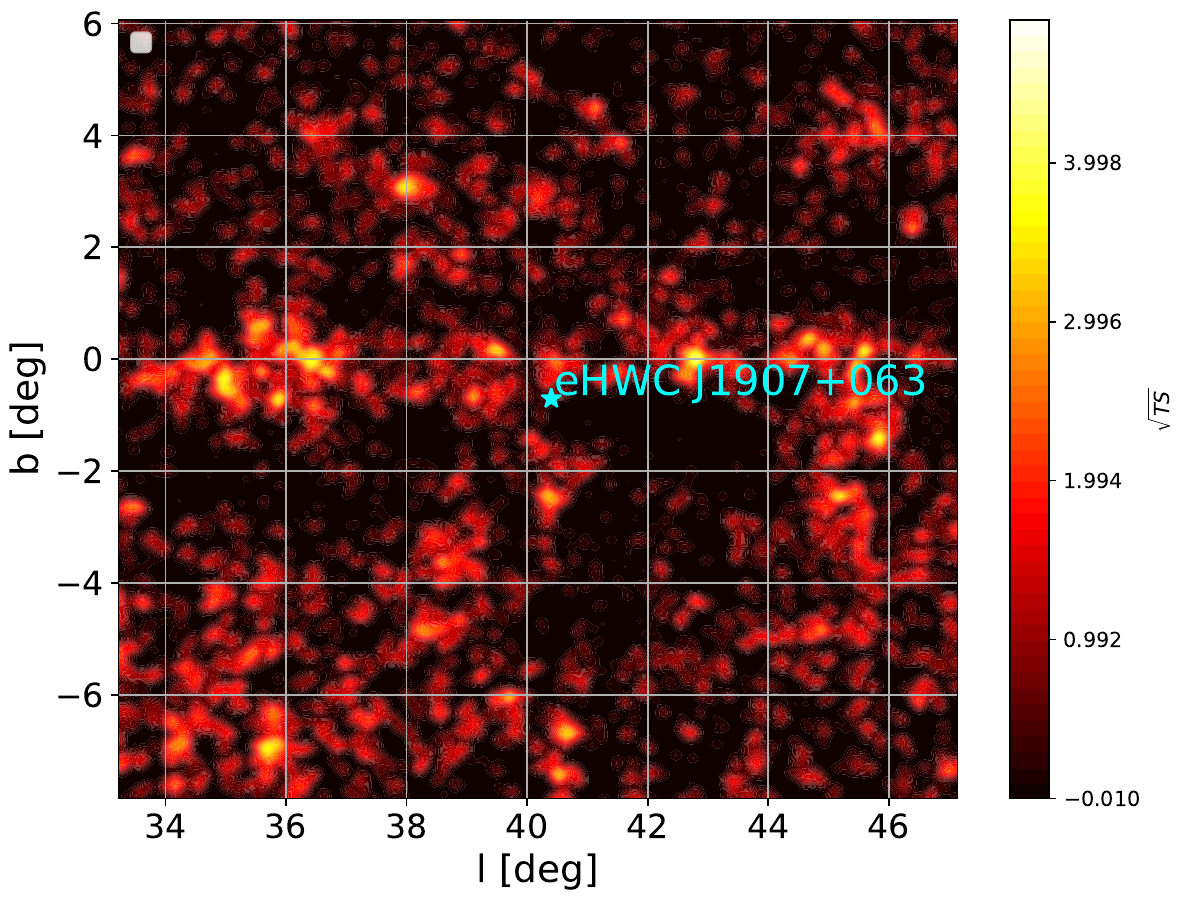}
\caption{$\sqrt{TS}$ map of the ROI analyzed in \fermilat data for the source eHWC J1907+063, after subtracting the best-fit model obtained as described in Sec.~\ref{sec:data}. This figure has been obtained using the background model IEM-\textit{4FGL}.}  
\label{fig:tsmaps_final1908}
\end{figure}

\begin{table*}
\begin{center}
\begin{tabular}{|c|c|c|c||c|c|c|}
\hline
 IEM  &   \textit{4FGL}  &  \textit{GC}  &  \textit{ALT1}   &   \textit{4FGL}  &  \textit{GC}  &  \textit{ALT1}  \\ 
\hline
\hline
 $l$ [deg] & $40.61\pm0.08$  & $40.61\pm0.15$  &  $40.71\pm0.17$  & $40.54\pm0.13$  & $40.59\pm0.14$  &  $40.50\pm0.11$ \\
 $b$ [deg] & $-0.62\pm0.08$  & $-0.47\pm0.08$  &  $-0.45\pm0.11$    & $-0.64\pm0.11$  & $-0.53\pm0.18$  &  $-0.44\pm0.11$ \\
 $\theta_{68}$ [deg] &  $0.77\pm0.05$ & $0.69\pm0.06$ & $0.65\pm0.09$     &  $0.73\pm0.10$ & $0.79\pm0.12$ & $0.79\pm0.12$ \\  
 $TS$($TS_{\rm{ext}}$)  & 91(55)  & 119(58)  &  217(53)               & 45(21)  & 60(29)  &  109(26) \\
\hline
\hline
\end{tabular}
\caption{Best-fit values for the position, extension and significance of detection for eHWC J1907+063, as found analyzing {\it Fermi}-LAT data in the energy range from 1 to 1000 GeV. These results are expressed with the longitude ($l$) and latitude ($b$), $68\%$ containment angle $\theta_{68}$, the $TS$ for the detection of the source and $TS$ of extension ($TS_{\rm{ext}}$). Each row block corresponds to  the three IEM considered in this paper while the three left (right) columns are for the standard (off-pulse) analysis.}
\label{tab:1908}
\end{center}
\end{table*}

The optimization process finds an extended source at the location of the source eHWC J1907+063 for each of the IEM models listed in Sec.~\ref{sec:data}. 
We test both a Gaussian and uniform disk templates finding that the former gives slightly larger detection significance. Thus we decide to provide the results for the radial Gaussian template spatial morphology. 
We report in Tab.~\ref{tab:1908} the results we obtain.
Fixing the IEM-\textit{4FGL}, IEM-\textit{GC} and IEM-\textit{ALT1} models we find an extension 
 $\theta_{68} = 0.71\pm0.05$/$0.69\pm0.06$/$0.65\pm0.09$ deg and a $TS_{\rm{EXT}}=55/58/53$\footnote{The $TS$ of extension is defined as $TS_{\rm{EXT}}=2 (Log(\mathcal{L_{\rm{PS}}})-Log(\mathcal{L_{\rm{EXT}}}))$ where $Log(\mathcal{L_{\rm{PS}}})$ is the likelihood found when using a point source template while $Log(\mathcal{L_{\rm{EXT}}})$ a Radial Gaussian template.}.
We also run the localization finding best-fit positions between the different IEMs that are compatible within the errors.

Since the pulsar J1907+060 is extremely bright in {\it Fermi}-LAT data, an imperfect modeling of the detector PSF could leave residuals around this source.
Therefore, a detection of a halo around the bright pulsar could be due to residuals left from imperfections of the modeling of the LAT PSF. 
We perform an off-pulse analysis, to see if we still detect an extended source\footnote{See this page for a complete description of this procedure \url{https://fermi.gsfc.nasa.gov/ssc/data/analysis/scitools/pulsar_gating_tutorial.html}}.
We select the data that are off from the peak of the pulsation of the pulsar (between 0.7 and 1.0), and we rerun the analysis.
In Tab.~\ref{tab:1908} the results we find with the three tested IEMs for the spatial extension and position are displayed.
We find similar values for the extension as found before, and with a lower $TS_{\rm{EXT}}=21/29/26$ for the IEM-\textit{4FGL}, IEM-\textit{GC} and IEM-\textit{ALT1}, respectively.
Also the position is compatible with the results we obtain with the standard analysis.
The lower values for the $TS_{\rm{EXT}}$ are due to the fewer photons available in the off-pulsed analysis with respect to the standard analysis.
Very recently Ref.~\cite{Li:2021wzt} performed an analysis towards this source but starting from 100 MeV. They also find an extended emission around this source with a similar extension and position as in our analysis.

\subsection{eHWC J2019+368}
\begin{figure*}[t]
\includegraphics[width=0.49\textwidth]{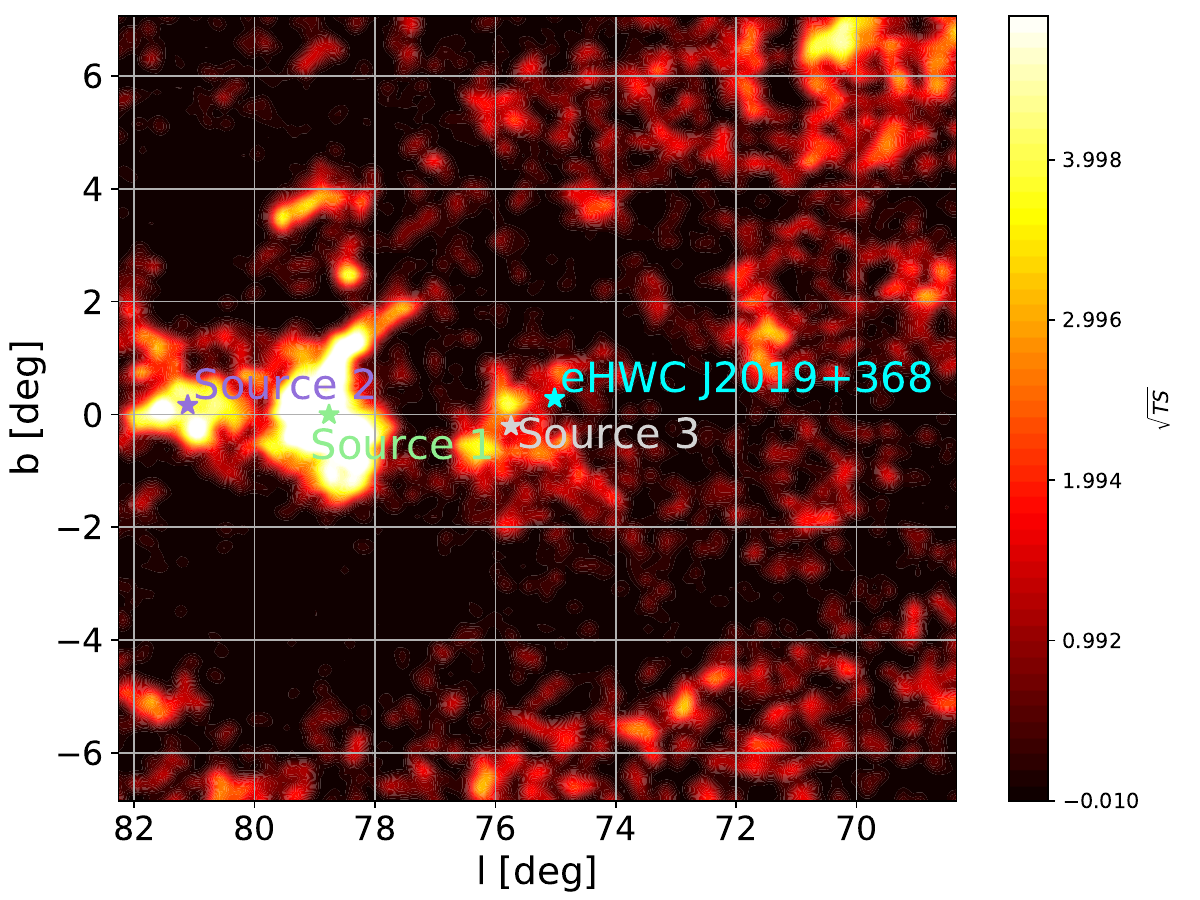}
\includegraphics[width=0.49\textwidth]{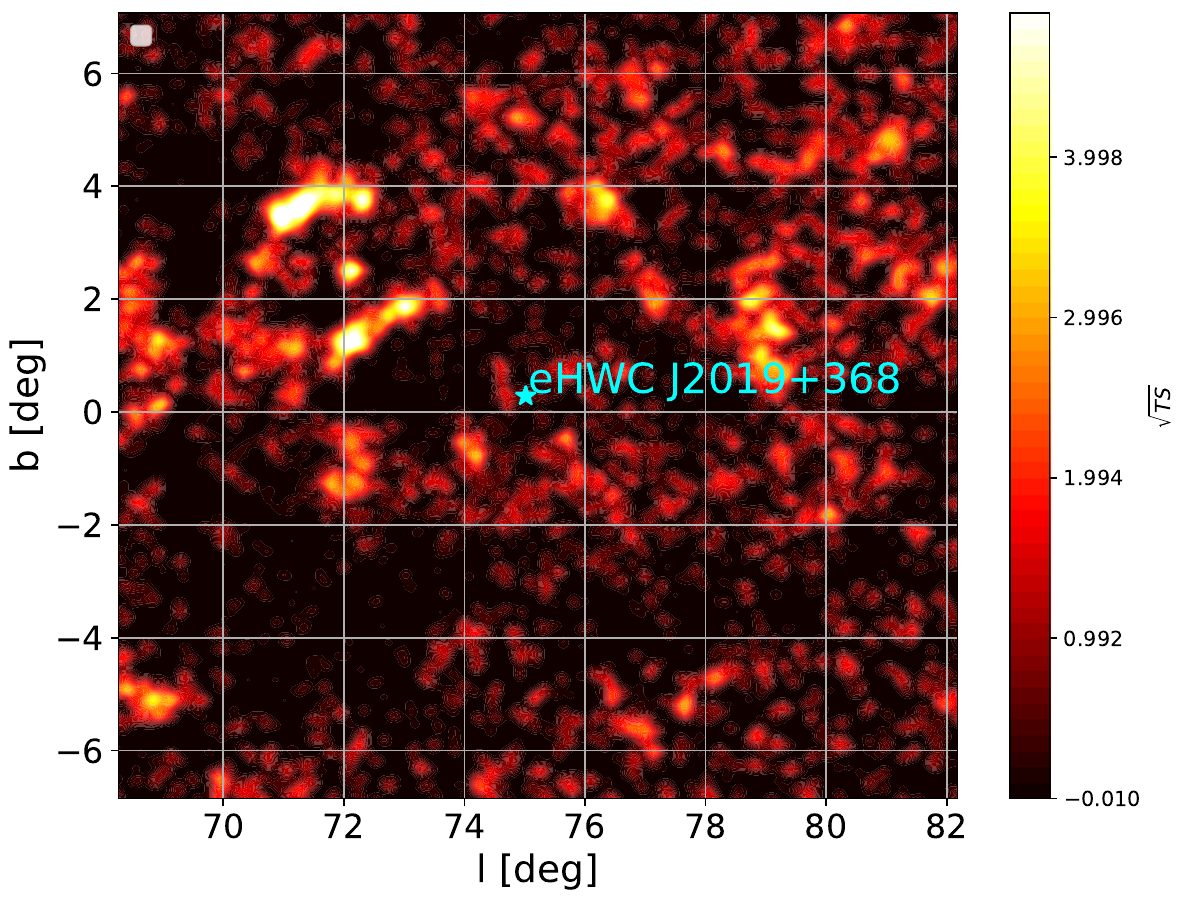}
\caption{$\sqrt{TS}$ map found in the energy range between 1-1000 GeV for the ROI considered around the source eHWC J2019+368. In the left panel we show the $\sqrt{TS}$ map that we find when we use the 4FGL sources, IEM and isotropic templates. We also display the position of the peaks that are re-absorbed when we add three additional extended sources. In the right panel we show the $\sqrt{TS}$ when  {\tt Source} 1, 2 and 3 are included in the background model.}  
\label{fig:tsmaps_final2019}
\end{figure*}

By performing the ROI optimization to eHWC J2019+368, we firstly find  significant residuals. 
We show the corresponding $\sqrt{TS}$ map using the IEM-\textit{4FGL} model in the left panel of Fig.~\ref{fig:tsmaps_final2019}. 
The residuals are mostly located on the Galactic plane, and are likely due to the fact that in the 4FGL catalog the Cygnus region is modeled by a simple 2D Gaussian with $3$ deg size, which poorly represents its complicated emission. 
We thus improve the 4FGL catalog model, that includes sources and interstellar emission, by searching new sources and quantifying their possible extension.
The optimization is done from 100 MeV, since the residuals could be due to un-modeled interstellar emission, which is brighter at lower energies.   
During this optimization we detect three new sources, labelled as \textit{Source 1, 2} and \textit{3}. 
They are found with a $TS$ of 1400, 740 and 250 and with an extension of $0.58$, $0.47$, $0.50$ deg using a Gaussian template, respectively.
We illustrate the position of these sources within the ROI in the left panel of Fig.~\ref{fig:tsmaps_final2019}.
Similar results are found using the IEM-\textit{GC} and IEM-\textit{ALT1}.

These new extended sources are probably associated with $\gamma$ rays produced from $\pi^0$ decays of freshly accelerated CRs interacting with gas atoms of the ISM.
Indeed,  their fluxes as a function of energy share a similar spectrum peaked at a few GeV, as shown in Fig.~\ref{fig:sedsources2019}. 
The flux (displayed as $E^2 dN/dE$) decreases significantly above a few GeV, meaning that these additional sources do not contribute significantly above 10 GeV. 
Also, we note that they all have roughly the same normalization at 3 GeV. 
The similar shape and normalization suggest a common origin for \textit{Source 1, 2} and \textit{3}.
When we include these new sources in the background model, we find that 
{\textit{Source 1, 2} and \textit{3} improve significantly the modeling of the eHWC J2019+368 ROI, as clearly visibile in the $\sqrt{TS}$ map displayed in right panel of Fig.~\ref{fig:tsmaps_final2019}.
When running the search of an ICS halo around eHWC J2019+368, we select only energies above 6 GeV with the spectrum of \textit{Source 1, 2} and \textit{3} fixed as found in the optimization process.
In this way we select energies where less residuals are expected, and we have more leverage to constrain a possible ICS halo.
We search for an extended source at the location of the pulsar J2021+3651 - associated with eHWC J2019+368 within a geometrical model of a radial disk or a Gaussian template. However, the presence of an extended source is not significant, with the $TS$ lower than 25.
We do not perform an off-pulsed analysis of this source since we do not find any evidence of an extended emission around the pulsar.
\begin{figure}[t]
\includegraphics[width=0.49\textwidth]{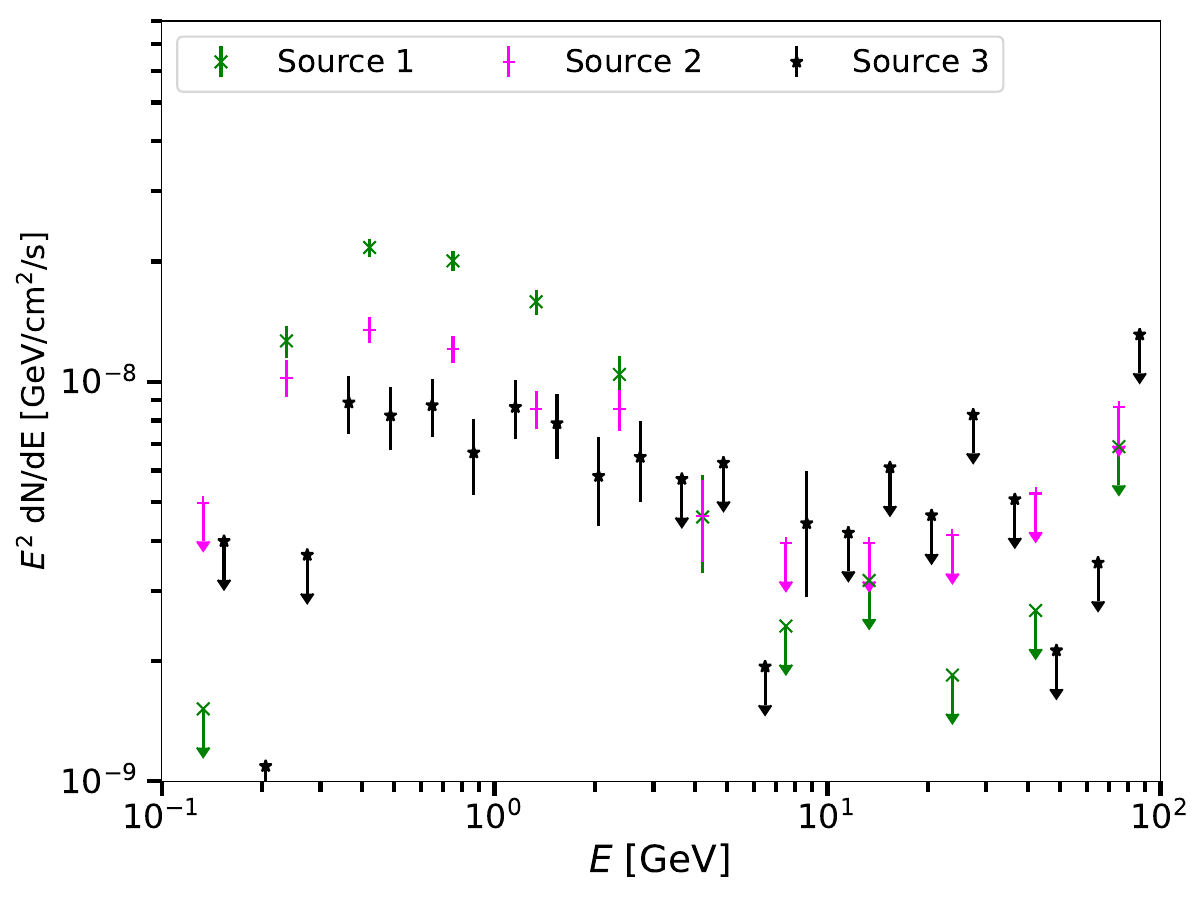}
\caption{Flux as a function of energy for the three extended sources found in the \fermilat ROI considered for the source eHWC J2019+368. Downward arrows stay for upper limits. }  
\label{fig:sedsources2019}
\end{figure}

\section{Results on the search for Inverse Compton halos}
\label{sec:ICShalo}
\begin{figure*}[t]
\includegraphics[width=0.49\textwidth]{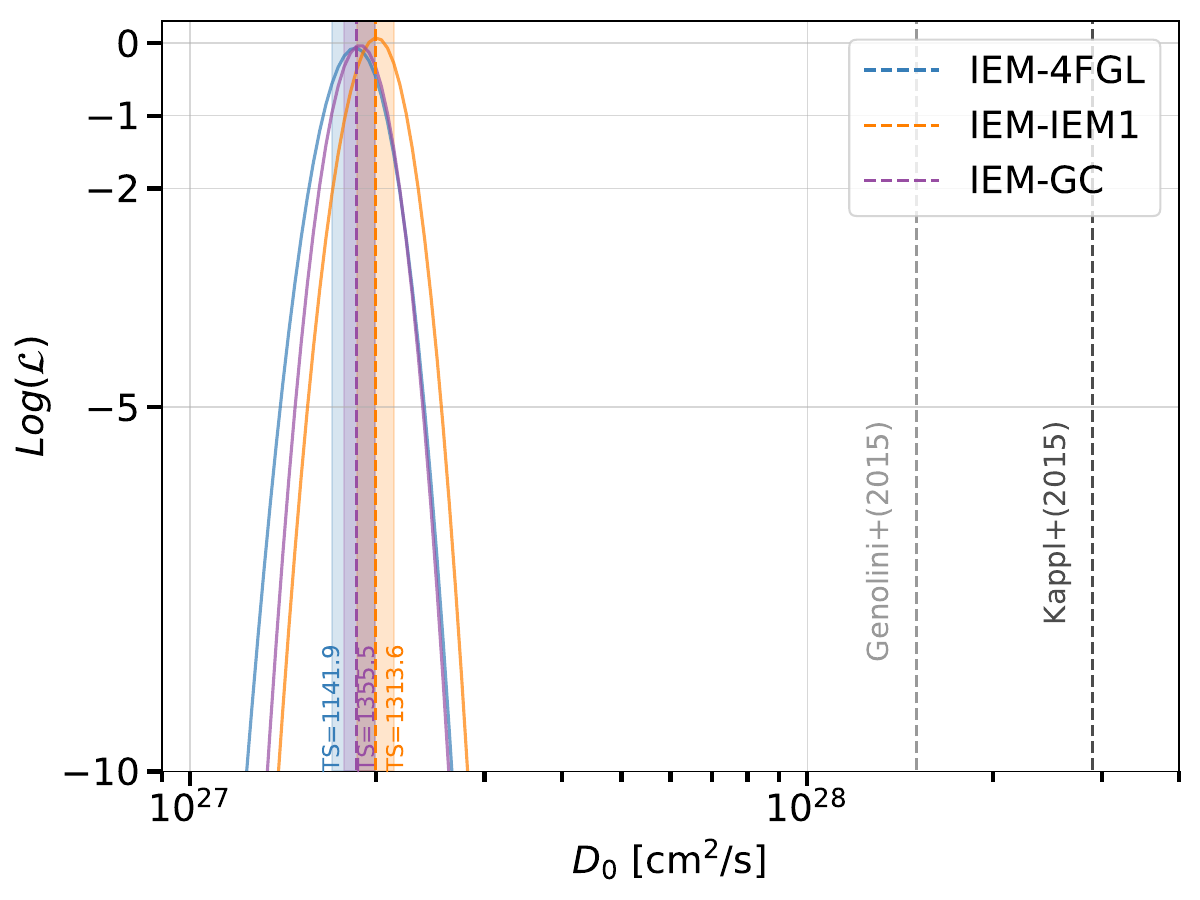}
\includegraphics[width=0.49\textwidth]{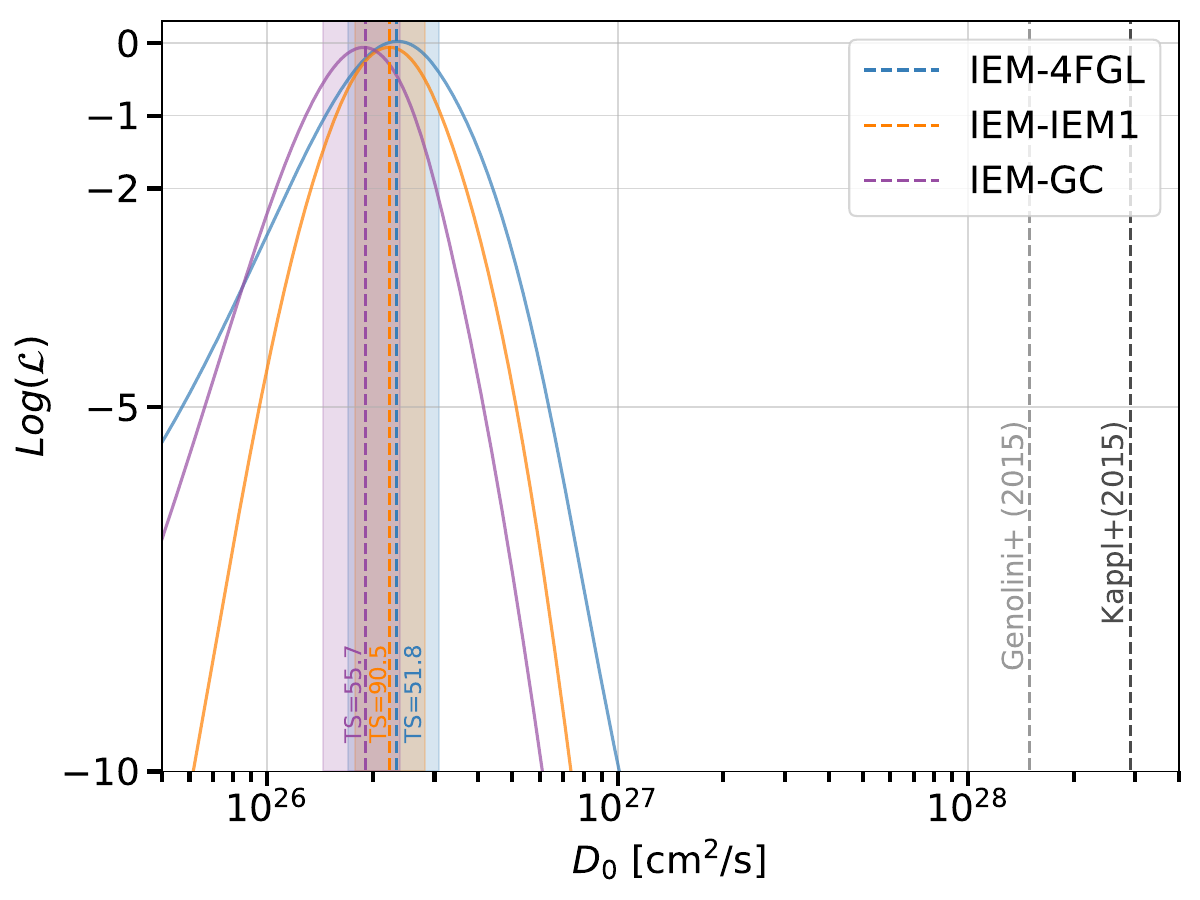}
\caption{Relative change in the $\log{\mathcal{L}}$ profile for eHWC1825-137 (left panel) 
and eHWC J1907+063 (right panel) as a function of the diffusion coefficient normalization $D_0$, and for three different IEMs. The dashed vertical lines show the $D_0$ values in correspondence to the maximum likelihood, whose respective TS values are reported in the plot. 
 Two reference values for $D_0$ found on Galactic scales are also reported 
 \cite{Kappl:2015bqa,Genolini:2019ewc}.}
\label{fig:likelihood}
\end{figure*}

 In this Section we report on the properties of the extended emission around the PWNe when studied within the physically-motivated ICS template.
Past analyses attempting to physically motivate the PWN/SNR gamma-ray flux and SED, assumed a disk or gaussian geometrical template with $e^-$ and/or $e^+$ source injection and propagation into the surrounding medium (e.g.~\cite{Principe:2019pql}).

Here, we generate ICS templates with the model explained in Sec.~\ref{sec:model} for different values of  $D_0$. 
Then, we find the value of  $D_0$, which gives the highest likelihood, i.e.~the best fit to the data, fitting {\it Fermi}-LAT data with a standard maximum likelihood analysis. 
The goal is to investigate the possible presence of a low diffusion zone around the PWNe, where $e^\pm$ would reside longer than if the diffusion was similar to the Galactic average.

In Fig.~\ref{fig:likelihood} we display the likelihood profile for eHWC J1907+063 and eHWC J1825-134 (the two sources for which we detected a significant extension with the geometric templates) as a function of $D_0$. 
The likelihood profiles  are peaked at $D_0\sim  2\times 10^{27}$ cm$^2$/s for eHWC J1825-134, and $D_0\sim  2\times 10^{26}$ cm$^2$/s for eHWC J1907+063. The position of the peaks does not change significantly for the analysis performed with the three different IEM models. This implies that our result for $D_0$ is robust with respect to systematics of the background modeling. 
The likelihood profile is much narrower for eHWC J1825-134 with respect to eHWC J1907+063, because the source is detected much more significantly.
The ICS model improves significantly the $TS$ of eHWC J1825-134, changing from a value of 846, obtained with the radial Gaussian modeling, to 1150 obtained within the ICS template with $D_0$ at its best fit value. 
Our $TS$ with the ICS template is also much higher than the value reported in the Ref.~\cite{Principe:2020zqe} where they found 1040 using a Gaussian template. 
 This is a very large $TS$ difference, considering that the ICS template and the Gaussian template have the same number of free parameters, that implies that the former performs much better than the second\footnote{The Gaussian template has the position, spectral index, normalization and size of extension while with the ICS template we substitute this latter parameter with the diffusion coefficient.}.
The result for eHWC J1825-134 has been obtained setting the center of the ICS template at the position of the pulsar. We find results compatible within $1\sigma$ errors when the ICS template is moved at the center of the $\gamma$-ray source detected at $E>10$ GeV (see Sec.~\ref{sec:1825opt}).
As for eHWC J1907+063, the $TS$ improves only mildly with the physical ICS template instead of the geometrical modeling. 
For example, with the IEM-{\it GC} the $TS$ changes from a value of 48 (radial Gaussian) to 57 (ICS template), with the ICS template  set  at the position of the pulsar. 

In Fig.~\ref{fig:likelihood} we also show that  the found $D_0$ value is not compatible with the commonly derived Galactic diffusion coefficient values \cite{Kappl:2015bqa,Genolini:2019ewc}.
The same analysis applied to eHWC J2019+368 does not provide any significant detection for an ICS halo, and the likelihood profile as a function of $D_0$ is almost flat. For this source, we are thus not able to provide a preferred value of $D_0$. This result is consistent with the non-detection of any extended emission when using the geometrical template during the ROI optimization.
The values for $D_0$ we find  for the sources eHWC J1907+063 and eHWC J1825-134 can be compared with the ones derived in \cite{Abeysekara:2017science,DiMauro:2019yvh,DiMauro:2019hwn} in the direction of different PWNe.
Specifically, Refs.~\cite{Abeysekara:2017science,DiMauro:2019yvh} found evidence for ICS halos around Geminga and Monogem in HAWC and {\it Fermi}-LAT data, with diffusion coefficient values spanning $D_0\sim  0.7-1.5 \times 10^{26}$ cm$^2$/s and $D_0\sim  4\times 10^{26}$ cm$^2$/s, respectively. Additionally, Ref.~\cite{DiMauro:2019hwn} found extended emission compatible with ICS halos around a sample of sources detected in the HESS survey of the Galactic plane with $D_0\sim 1-10 \times 10^{26}$ cm$^2$/s.

From the analysis of the ICS template, we also find the \fermilat SED data points.  They are reported in Figs.~\ref{fig:SED1}, \ref{fig:SED2} and \ref{fig:SED3}, when 
fixing the diffusion coefficient to the best-fit value obtained from the maximum likelihood analysis of {\it Fermi}-LAT data.  
The results are stable with variations  in the IEM, specifically using IEM-\textit{GC} and IEM-\textit{ALT1}. 
 In each figure, together with the data obtained with our analysis of {\it Fermi}-LAT data, we also display the measurements reported by the HAWC Collaboration \cite{Abeysekara:2019gov}.
Since for the source eHWC J2019+368 we do not report any detection of a ICS halo, we fix $D_0$ to $\sim  3\times 10^{26}$ cm$^2$/s and we find upper limits for the flux. 
The value we choose is representative of the recent detections of ICS halos around pulsars  \cite{Abeysekara:2017science,DiMauro:2019yvh,DiMauro:2019hwn}. However, the results for the upper limits are not significantly affected by this choice.
 We remind the reader that, we fix the diffusion coefficient, that mainly modifies the spatial extension, by performing the pixel-by-pixel and energy bin maximum likelihood analysis presented in this section. Instead, the parameters, such as $\gamma_e$ and $\eta$, that modify the spectral part of the model, are derived by fitting the {\it Fermi}-LAT and HAWC SED measurements.

\begin{figure}[t]
\includegraphics[width=0.48\textwidth]{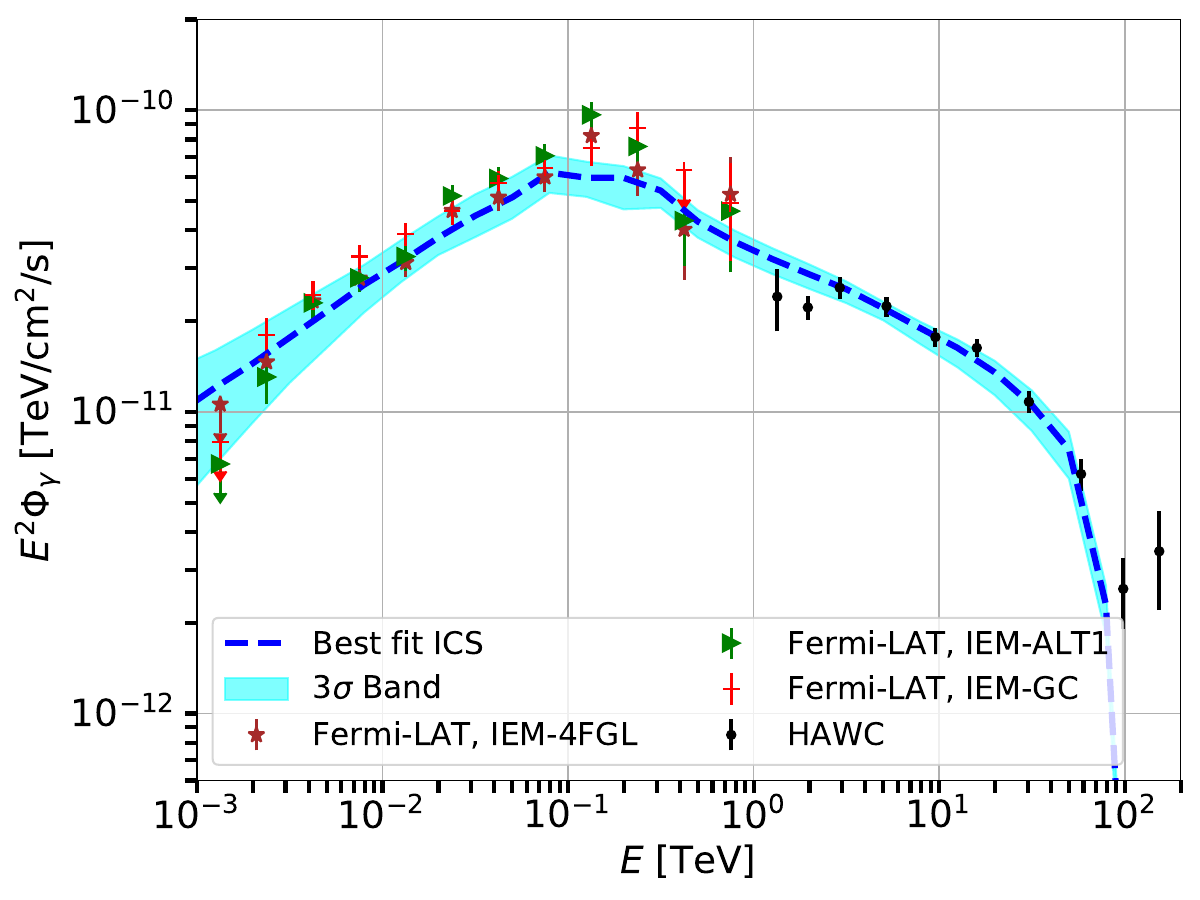}
\caption{Flux as a function of energy found for the source eHWC J1825-134 with the ICS template generated at the best-fit value of $D_0$ found with our analysis. We show the data found when using three different IEMs. Together with the flux data we also show the best-fit and the $3\sigma$ band for the ICS theoretical predictions found by fitting the flux data.}
\label{fig:SED1}
\end{figure}
\begin{figure}[t]
\includegraphics[width=0.48\textwidth]{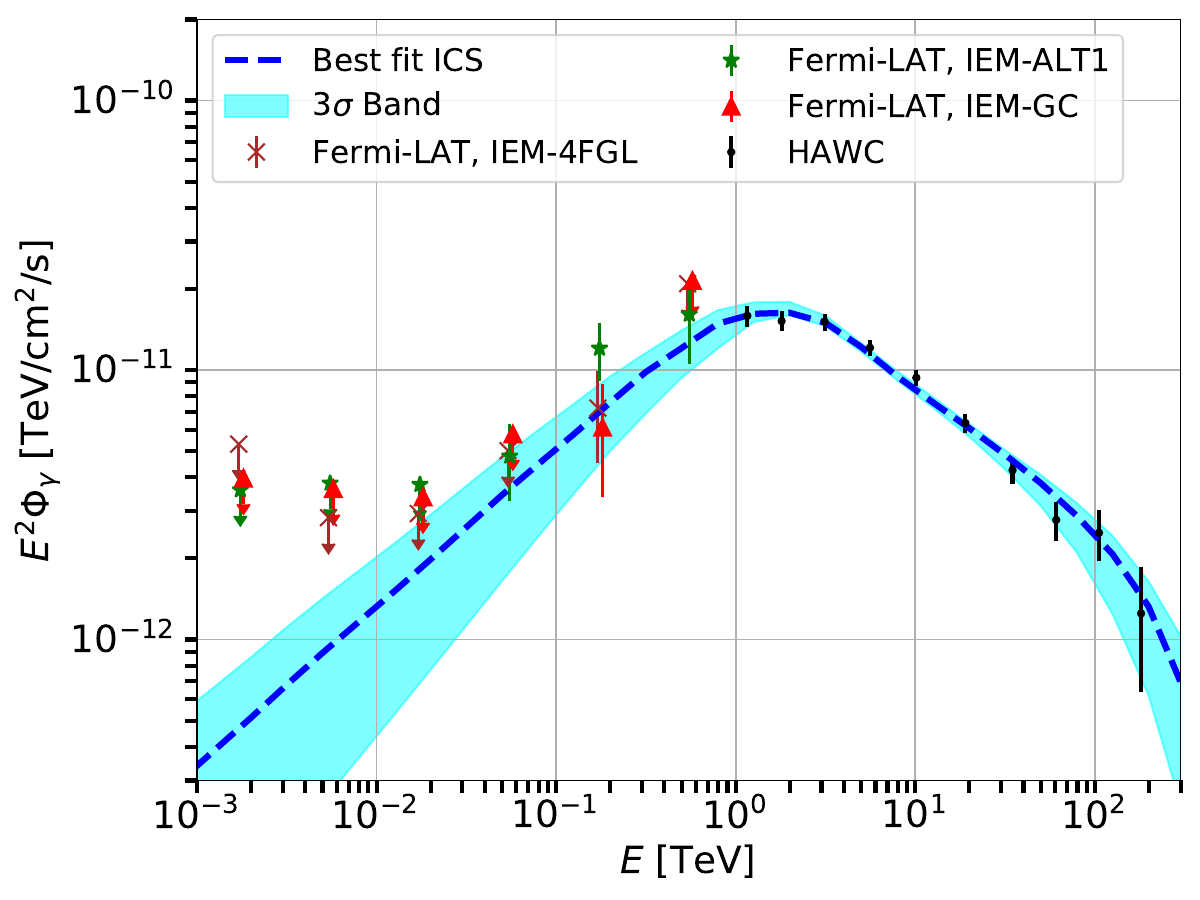}
\caption{Same as Fig.~\ref{fig:SED1} for the source eHWC J1907+063.}
\label{fig:SED2}
\end{figure}
\begin{figure}[t]
\includegraphics[width=0.48\textwidth]{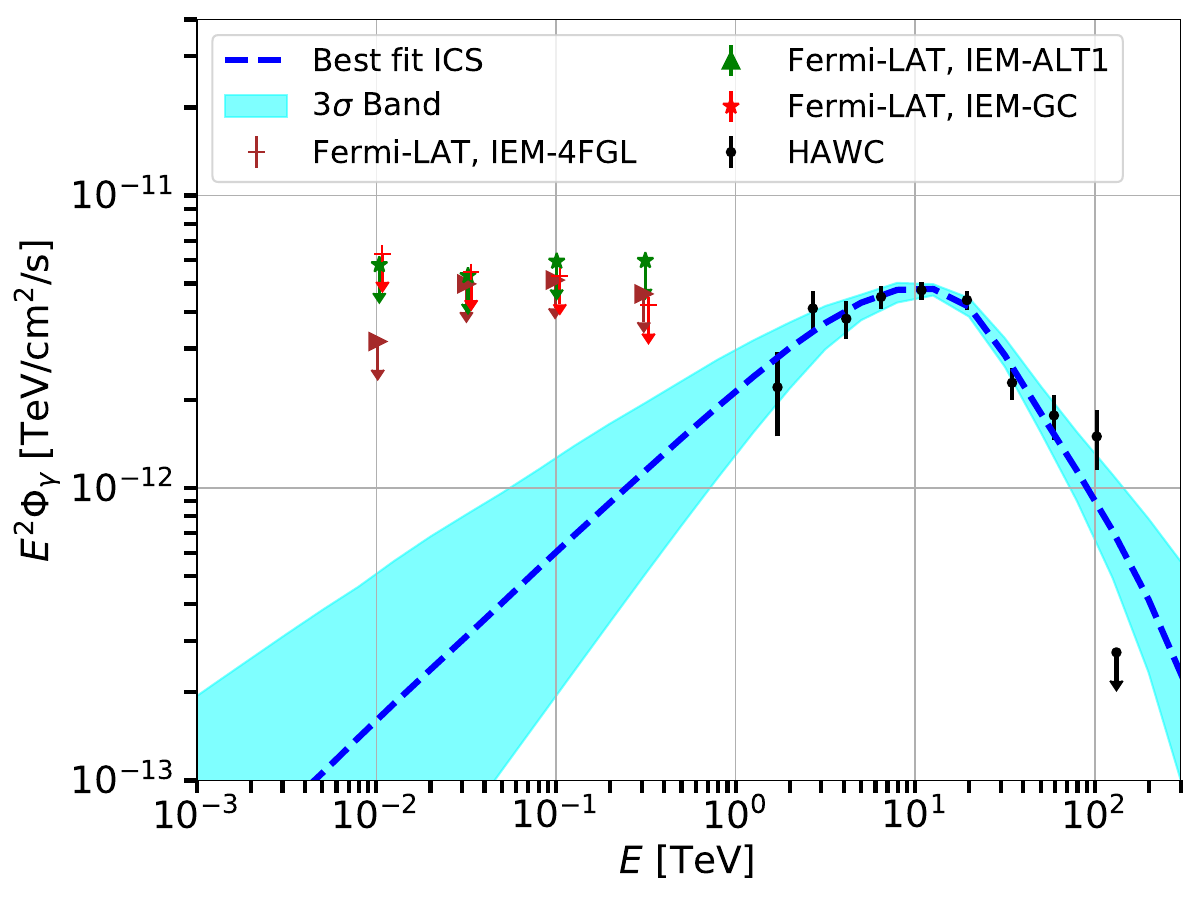}
\caption{Same as Fig.~\ref{fig:SED1} for the source eHWC J2019+368.}
\label{fig:SED3}
\end{figure}

\section{Discussion}

In order to understand the properties of the $e^\pm$ parent population, we study the $\gamma$-ray SED obtained with the analysis on \fermilat data together with the HAWC SED.
In this procedure we use consistently the same model of $\gamma$-ray flux for ICS.

From Figs.~\ref{fig:SED1}, Fig.~\ref{fig:SED2}, Fig.~\ref{fig:SED3}, we notice that the best fits to the SED data as a function of the energy have a bumpy shape for all the sources. The peak of the $\gamma$-ray flux is located at 0.1/1/10 TeV for eHWC J1825-134, eHWC J1907+063, eHWC J2019+368, respectively. It is very likely 
the result of the different energy losses suffered by $e^{\pm}$ injected by the PWNe, and traveling in the surroundings of the source.
Moreover, all the three spectra hint at a cutoff at energies above a few tens of TeV. 
It could be an intrinsic cutoff in the injection spectrum of $e^\pm$ by the PWNe. 
However, this feature is also compatible with the softening of the flux caused by the propagation of $e^{\pm}$ in the Galaxy. 
Indeed, $e^{\pm}$ with energy above 500 TeV loose energy very quickly, and the probability to produce $\gamma$ rays at such high energies is very low. A rough estimate of the maximum energy of a $e^\pm$ produced by the source eHWC J1825-134 can be performed starting from  the inverse of its 
energy loss rate $\sim 1/(b_0 t)$,  where  $t$ is the time on the pulsar era at which the $e^\pm$ is emitted. 
We can approximate the energy losses at these $e^{\pm}$ energies as $b_0\sim 3\cdot 10^{-17}$ GeV$^{-1}$ s$^{-1}$, and consider a time $t\sim \tau_0 = 12$ kyr before which most of the energy of the pulsar is emitted.  Using these approximations, we find that the maximum $e^{\pm}$ energy is about 80 TeV, that is roughly the energy above which hint at a cutoff in the spectrum in Fig.~\ref{fig:SED1}, \ref{fig:SED2} and \ref{fig:SED3} is observed.
We see in Fig.~\ref{fig:SED1} that the cutoff for the ICS flux in our best-fit model for eHWC J1825-134 does not reproduce well the two highest energy data points above 100 TeV, which can still be explained lowering the value we use for the energy losses below $10^{-16}$ GeV/s. Instead, the highest energy point is difficult to reconcile with our model and could be the hint of an additional component of CRs emitted by the source. 
These two data points could be due to an un-modeled hadronic emission.
Our model is not compatible with the HAWC upper limit found for the source eHWC J2019+368 in the highest energy data point (see Fig.~\ref{fig:SED3}). This could be due to a more stringent cutoff required in the injection spectrum of $e^{\pm}$ from this source.

Under the hypothesis that the ICS halo is generated by $e^{\pm}$ emitted by the pulsar, the $\gamma$-ray flux as a function of energy can be used to constrain their injection spectrum, as done in Refs.~\cite{DiMauro:2019yvh,DiMauro:2019hwn}.
We perform a combined fit to the {\it Fermi}-LAT and HAWC SED data points minimizing the $\chi^2$ against the parameters of the model that change the spectral part of the ICS $\gamma$-ray flux:  spectral index for the injection of $e^{\pm}$ $\gamma_e$, the efficiency $\eta$ for the conversion of pulsars spin-down luminosity into $e^{\pm}$ pairs, the energy cutoff $E_c$, and the normalization factor $b_0$ of the energy losses suffered by these particles after being produced by the PWN. 
We show in Fig.~\ref{fig:SED1}, \ref{fig:SED2} and \ref{fig:SED3} the best-fit predictions for the ICS emission connected to 
the $e^{\pm}$ source spectrum, which are compatible with the flux data within $3\sigma$, after minimizing against  $\gamma_e$, $\eta$, $E_c$ and $b_0$.
\begin{table*}
\begin{center}
\begin{tabular}{|c|c|c|c|c|}
\hline
\hline
Source  &   $\gamma_e$  & $\eta$  &  $b_0$ [GeV/s]   &  $E_c$ [TeV] \\ 
\hline
eHWC J1825-134 &  $1.95\pm0.05$  &  $4.4\pm0.6$  &  $(2.0\pm0.5)\times 10^{-16}$  & $>500$  \\
\hline
eHWC J1907+063 &  $1.80\pm0.20$  &  $0.10\pm0.5$ &  $(6.0\pm1.0)\times 10^{-17}$  &  $>300$  \\
\hline
eHWC J2019+368 &  $1.90\pm0.20$  &  $0.008\pm0.004$  &  $(2.0\pm0.5)\times 10^{-17}$  &  $>300$  \\
\hline
\hline
\end{tabular}
\caption{Best-fit values for the parameters $\gamma_e$, $\eta$ and $b_0$ and the $3\sigma$ lower limit for $E_c$ found by fitting the 
$\gamma$-ray flux data shown in Fig.~\ref{fig:SED1}, \ref{fig:SED2} and \ref{fig:SED3}, within the IEM-\textit{4FGL}. }
\label{tab:resultspar}
\end{center}
\end{table*}
The results of the fits found within the IEM-\textit{4FGL} are reported in Tab.~\ref{tab:resultspar}. We find very similar best-fit values using 
IEM-\textit{GC} and IEM-\textit{ALT1}. 
The $e^\pm$ injection spectral indexes are found similar for the three sources and around $\gamma_e \simeq 1.90$;  instead the efficiency varies from $440\%$ for eHWC J1825-134 to much smaller values for the other two PWNe. The extremely low value for the eHWC J2019+368 efficiency is indeed meaningless, since it extrapolates the HAWC data to low energy, not affecting the \fermilat upper limits. 
For the source eHWC J1825-134 we find an efficiency which is larger than $100\%$. 
Although this result might be difficult to reconcile with the pulsar's energetics, 
this is consistent with what was found in Ref.~\cite{Sudoh:2021avj}
(Fig. 5 right panels) for a similar value of the
diffusion coefficient that we find in this paper. 
Moreover, Ref.~\cite{Principe:2019pql} found an
efficiency of about $50\%$ by fitting data from HESS and
Fermi-LAT which are lower by a factor among 2
and 7 in the energy range between a few GeV to a few TeV with respect
to our {\it Fermi}-LAT and HAWC data.

Also the best fit value for $b_0$ changes significantly among the three sources, going from $2.0 \times 10^{-16}$ GeV/s for eHWC J1825-134 to few times  smaller values for the other two sources.
This behaviour traces  the position of the flux peak, which appears around 0.1 TeV for eHWC J1825-134, about few TeV for eHWC J1907+063 and 
above 10 TeV for eHWC J2019+368. 
The intensity of the energy losses in eHWC J1825-134, $b_0 \sim 2\times 10^{-16}$ GeV/s, is found for $e^\pm$ energies  $E_e=[10,10^4]$ GeV, ICS off the local ISRF spectrum from \cite{Porter_2006} and synchrotron radiating off a magnetic field of about 5 $\mu$G.
On other hand, a lower value of $b_0 \sim 6\times 10^{-17}$  GeV/s and $2\times 10^{-17}$  GeV/s, as found for eHWC J1907+063 and eHWC J2019+368 respectively, is in principle compatible with an ISRF smaller by a factor of 2 and 3 from the local model in \cite{Porter_2006}  and a magnetic field of 4 and 3 $\mu$G, respectively. 
Differences of a factor of 2-3 in the density of the starlight and infrared components of the ISRF or in the value of the Galactic magnetic field with respect to the local values are viable, and could therefore explain the energy losses rate derived from the $\gamma$ rays in this analysis.
Indeed, the difference between the local and the Galactic ISRFs and magnetic fields could reach a factor of roughly 10 \cite{Porter_2006}.

As a final result, we derive a $3\sigma$ lower bound on the cutoff energy $E_c$ of the $e^{\pm}$ injection spectrum,  set at  $\sim 3-500$ TeV for three sources.
Such a lower limit implies that these three PWNe very likely accelerate $e^{\pm}$ up to PeV energies.
Moreover, since the value of the cutoff energy is not well constrained, the softening of the $\gamma$-ray SEDs is probably due to energy losses rather than intrinsic cutoff in the $e^{\pm}$ injection spectra.
The HAWC observation of photons from these sources up to 100 TeV has important consequences for the acceleration of $e^{\pm}$ from PWNe. Indeed, by looking to Fig.~\ref{fig:icpower}, one can notice that such very-high-energy photons are mostly produced from $e^{\pm}$ at about 200-400 TeV. However, PeV electrons could produce at least about 10\% of these photons (see orange-red regions), thus justifying a leptonic origin of the observed $\gamma$-ray flux from eHWC J1825-134, eHWC J1907+063 and eHWC J2019+368. 

Multiwavelength campaigns have been performed to detect the PWN around the pulsars associated with eHWC J1825-134 and eHWC J1907+063 \cite{1996ApJ...466..938F,Gaensler:2002wq,Pavlov:2007av,Uchiyama:2008bd,Pandel:2015lfz,Liu:2019vlz,Duvidovich:2019ldj,2020MNRAS.491.5732D}.
Ref.~\cite{Uchiyama:2008bd} used Suzaku observations in a region $19\times19$ arcmin$^2$ around eHWC J1825-134 and reported an upper limit for the PWN flux of $5.4\cdot 10^{-9}$ GeV/cm$^2$/s in the energy range between $0.2-12$ keV.
Instead, Ref.~\cite{Pandel:2015lfz} published XMM-Newton observations of an ROI of $45\times45$ arcmin$^2$ around eHWC J1907+063 finding an upper limit for the PWN flux of $4.4\cdot 10^{-9}$ GeV/cm$^2$/s in the energy range between $1-10$ keV.
We use these upper limits to constrain the magnetic field around the pulsars associated with those sources. 
In particular we take the best-fit model we derived from the fit to $\gamma$-ray data. 
We use the same $e^{\pm}$ population that produce the $\gamma$-ray emission for ICS and we calculate the Synchrotron radiation they produce due to the PWN magnetic field as described in Ref.~\cite{DiMauro:2019yvh}. We perform the calculation for the field of view of X-ray observations and we vary the value of the magnetic field until we reach a flux equivalent to the measured upper limit. We find upper limits for the magnetic field strengths of 11 and 13 $\mu$G respectively for eHWC J1825-134 and eHWC J1907+063 that are compatible with the strength obtained by fitting {\it Fermi}-LAT and HAWC $\gamma$-ray data.

\section{Conclusions}
\label{sec: conclusions}
Following the discovery of three $\gamma$-ray sources by HAWC at energies E$>100$~TeV \cite{Abeysekara:2019gov}, we  
investigate the presence of extended $\gamma$-ray emission in \fermilat data around eHWC J1825-134, eHWC J1907+063 and eHWC J2019+368 PWNe. We study each source with an ICS template, where the extension of the $\gamma$-ray emission is implicitly given by the $e^\pm$ produced by the PWN, then propagating and losing energy around the source and in the Galaxy. Our main results on the analysis of \fermilat data can be  summarized as follows.
\\
\begin{itemize}
\item We find an extended emission around eHWC J1825-134 at high significance, with $\theta_{68} = 1.00^{+0.05}_{-0.07}$ deg. 
The result is robust against a number of systematics checks and compatible with previous estimates for this source \cite{Principe:2019pql}
\item We use the ROI optimization process to find an extended source at the location of the source eHWC J1907+063 with an extension  $\theta_{68} = 0.71\pm0.10$ deg,  which is confirmed after different IEMs and an off-pulse analysis.
\item  In the ROI optimization process around eHWC J2019+368 we find significant residuals, which lead us to the identification of three new sources around it. Even if we  include these new sources in the background model, the presence of an extended source is not significant. 
\item We find that the peak of the $\gamma$-ray flux is located at 0.1/1/10 TeV for eHWC J1825, eHWC J1907+063, eHWC J2019+368, and is understood as the effect of the different energy losses suffered by $e^{\pm}$ injected by the PWNe.
\item The ICS template fits the data for a diffusion coefficient value which is significantly lower than the average Galactic one. 
The likelihood profile is peaked at $D_0\sim  2\times 10^{27}$ cm$^2$/s for eHWC J1825-134, and $D_0\sim  2\times 10^{26}$ cm$^2$/s for eHWC J1907+063. The result is robust with respect to systematics of the background modeling. 
\end{itemize}
In order to understand the properties of the $e^\pm$ parent population, we study the $\gamma$-ray SED obtained with the present analysis on  \fermilat data  together with the HAWC one.
\\
We minimize against the spectral index for the injection of $e^{\pm}$ $\gamma_e$, the efficiency 
$\eta$ for the conversion of pulsars spin-down luminosity into $e^{\pm}$ pairs, and the normalization factor $b_0$ of the energy losses suffered by these particles after being produced by the PWN. We also set lower bounds on the energy of a possible cut-off in the $e^{\pm}$ injection spectrum, 
and argue that the softening of the $\gamma$-ray SED above a few TeV is compatible with energy losses suffered by $e^{\pm}$ for synchrotron emission and ICS.
Our results corroborate the existence of extended GeV-TeV $\gamma$-ray emission around PWNe, and connect indissolubly the radiation at its highest energies with $e^\pm$ populations slowly diffusing around PWN, even if we remind that other processes such as advection could play a role at GeV energies. These discoveries add a tile on the road of understanding the highest energy phenomena occurring in our Galaxy, and offer fundamental physics clues on the amount of antimatter produced in few kpc around our planet.

\begin{acknowledgments}
The work of FD and partially of SM has been supported by the "Departments of Excellence 2018 - 2022" Grant awarded by
the Italian Ministry of Education, University and Research (MIUR) (L. 232/2016).
FD acknowledges financial contribution from the agreement ASI-INAF
n.2017-14-H.0.
MDM acknowledges support from the Fellini Fellowship for Innovation at INFN, funded by the European Unions Horizon 2020 research programm under the Marie Sklodowska-Curie Cofund Action, grant agreement no. 754496

The \textit{Fermi} LAT Collaboration acknowledges generous ongoing support
from a number of agencies and institutes that have supported both the
development and the operation of the LAT as well as scientific data analysis.
These include the National Aeronautics and Space Administration and the
Department of Energy in the United States, the Commissariat \`a l'Energie Atomique
and the Centre National de la Recherche Scientifique / Institut National de Physique
Nucl\'eaire et de Physique des Particules in France, the Agenzia Spaziale Italiana
and the Istituto Nazionale di Fisica Nucleare in Italy, the Ministry of Education,
Culture, Sports, Science and Technology (MEXT), High Energy Accelerator Research
Organization (KEK) and Japan Aerospace Exploration Agency (JAXA) in Japan, and
the K.~A.~Wallenberg Foundation, the Swedish Research Council and the
Swedish National Space Board in Sweden.
 
Additional support for science analysis during the operations phase is gratefully
acknowledged from the Istituto Nazionale di Astrofisica in Italy and the Centre
National d'\'Etudes Spatiales in France. This work performed in part under DOE
Contract DE-AC02-76SF00515.

\end{acknowledgments}

\bibliography{paper}

\appendix

\section{Pulsar proper motion}\label{app:pmotion}

\begin{figure}[t]
\includegraphics[width=0.5\textwidth]{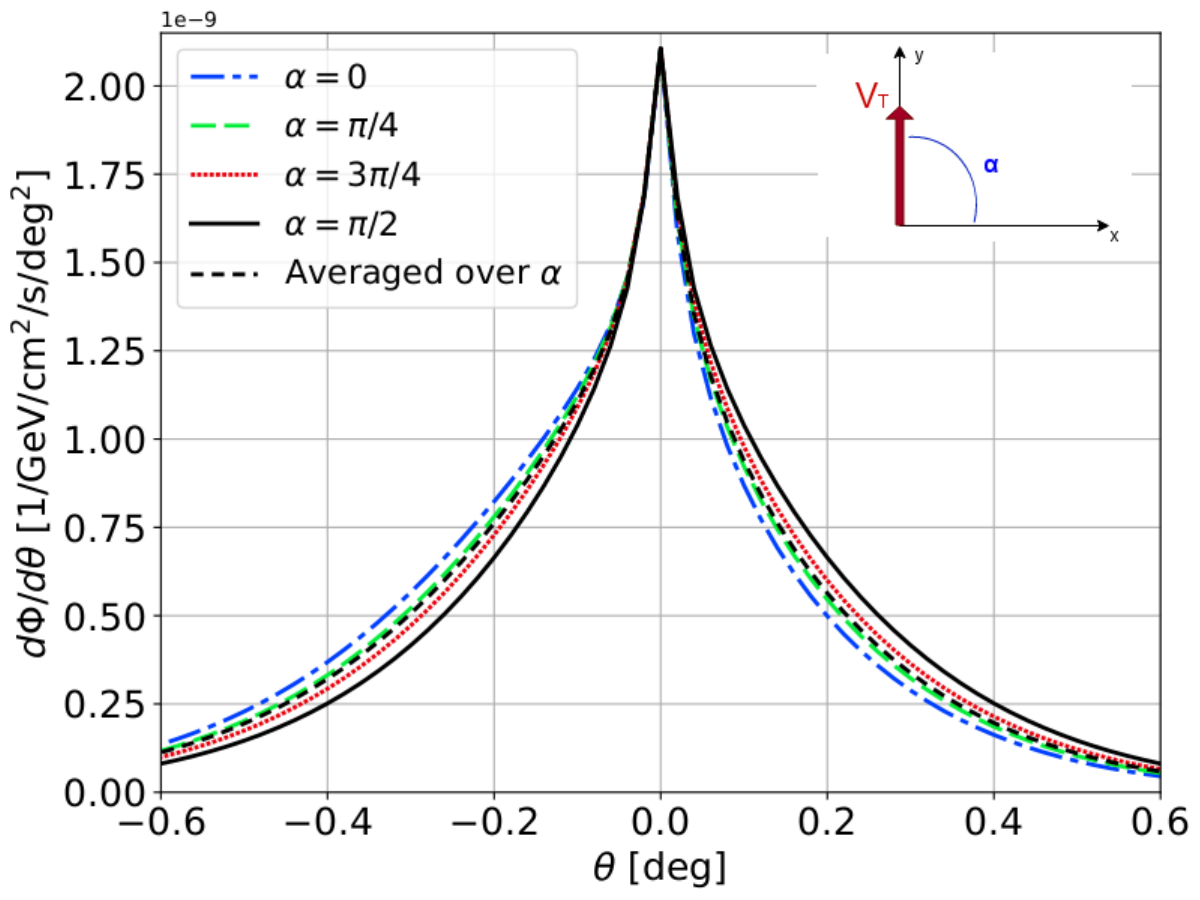} 
\caption{Study of the effect of the proper motion on the J1826-1334 surface brightness at 5 GeV, as a function of the observed angular distance $\theta$ from the centre of the pulsar. We assume $v_T=200$~km/s, and show surface brightness for different angles $\alpha$, defined with respect to the direction of $\vec{v}_T$ (fixed at $\alpha=\pi/2$).
The solid black curve represents the minimal distortion of the surface brightness obtained for $\alpha=\pi/2$ (i.e., $\theta$ changes along the $x$ axis and thus perpendicular to $\vec{v}_T$), while the other lines are for intermediate angles $\alpha=0 $ (blue), $\pi/4$ (green), and $3\pi/4$ (red). The dashed black line indicates the value averaged over $\alpha$.}  
\label{fig:propermotion}
\end{figure}
The pulsar proper motion was demonstrate to shape the morphology of the observed ICS emission at GeV energies in Refs.~\cite{DiMauro:2019yvh,DiMauro:2019hwn}.
In particular, at fixed distance and age of the source, the effect of the proper motion on the $\gamma$-ray morphology is governed by the pulsar transverse velocity $\vec{v}_T$, which is defined as the projection of the velocity of the source on a $xy$ plane perpendicular to the line of sight. 
In this work, the effect of the proper motion on the sample of the three sources (see Sec.~\ref{sec:sources}) will not be considered, given their age and distance to us. 
To motivate our choice, we study the possible effect of proper motion by simulating an ICS emission emitted from the brightest source in our sample, J1826-1334 (see Tab.~\ref{tab:sample}).
We assume a $\gamma$-ray energy of $E_\gamma=5$~GeV, which is close to the lower end of our energy range, because the effect of proper motion is larger at lower energies \cite{DiMauro:2019yvh,DiMauro:2019hwn}.

The geometry of this case of study is illustrated in the inset of Fig.~\ref{fig:propermotion}. 
The $xy$ plane is perpendicular to the line of sight and we artificially set the pulsar motion with a transverse velocity $\vec{v}_T$ aligned on the $y$ axis.
We assume that $v_T= |\vec{v}_T|=200$~km/s, which represents a rough average for Galactic pulsars' proper velocity \cite{Faherty:2007}. 
We also introduce  the opening angle $\alpha$  with respect to the $y$-axis.
We then compute the surface brightness $d\Phi/d\theta$ calculated for different angular distances $\theta$ between the direction that points towards the center of the source and the line of sight. 
We calculate the surface brightness by choosing different angles $\alpha$ with respect to the direction of $\vec{v}_T$.
The results are reported in Fig.~\ref{fig:propermotion}, and predict the effect of the proper motion on the J1825-137 surface brightness. 
As expected, the distortion in the surface brightness is maximal if it is calculated in the direction of $\vec{v}_T$ (i.e., $\alpha=0$). 
However, the effect is at most of the order of $35\%$ looking at the difference of the flux between $\theta = \pm0.2$ deg.
For any other direction, the distortion predicted in the $\gamma$-ray flux is negligible and smaller than the typical uncertainties in the measured source extension. 
We conclude that the effect of the pulsar proper motion on the observed surface brightness, that is averaged over $\alpha$, is negligible. 
This is understood in terms of the young age and of the distance of this source (see e.g. Fig.~5 in Ref.~\cite{DiMauro:2019hwn}). 
Similar conclusions are valid for the other two sources in our sample.

\section{eHWC J1825-134 and the pulsar J1826-1334}\label{sec:sourceJ1825} 
This source is one of the most studied very-high energy PWN, given its high luminosity, peculiar morphology and physical extension, which has a diameter of about $100$~pc (assuming a $4$~kpc distances \cite{Aharonian:2005kj}). 
It is identified as the PWN associated with the pulsar J1826-1334 (also known as B1823-13), a young ($T = 21$~kyr) and high spin-down pulsar ($\dot{E}=2.8\times 10^{36}$~erg/s).  
 
eHWC J1825-134 is the most significant and extended source reported by HAWC \cite{Abeysekara:2019gov}, with a $\sqrt{TS}=14.5$ ($\sqrt{TS}=7.33$) at energies larger than $56$~TeV ($100$~TeV). 
The pulsar J1826-1334 is found $0.26$~deg away from the center of the HAWC emission. Another pulsar of the ATNF catalog, J1826-1256, is found at $0.45$~deg. 
The physical extent of the HAWC emission, if associated with J1826-1334 at a distance to the Earth of $3.61$~kpc, is of $22.1$~pc. It corresponds to an angular extension of $0.36\pm 0.05$~deg at energies larger than $56$~TeV when fitting the source with a Gaussian morphology. 
The spectral energy distribution (SED) of the $\gamma$-ray emission is better fitted by a power-law with an exponential cutoff at $61\pm12$~TeV. If  these photons are interpreted as coming from ICS emission (Fig.~\ref{fig:icpower}), it implies the existence of $e^\pm$ accelerated to energies higher than $100$~TeV. 
The distance of this source reported in the ATNF catalog is 3.61 (3.93) kpc with the electron-density model \cite{2017ApJ...835...29Y} (\cite{Cordes:2002wz}).
This difference is not going to affect significantly any of the conclusion of this paper.

The presence of GeV $\gamma$-ray emission around the pulsar J1826-1334 was first claimed using \fermilat data by Ref.~\cite{Grondin:2011kw}, which found an extended nebula  of $0.56\pm 0.07$~deg in the energy range $1-100$~GeV (assuming a geometrical Gaussian model for the emission). 
This source has been then included in the \fermilat catalog of extended sources in the $10$~GeV-$1$~TeV energy band \cite{Ackermann:2017hri}. 
A recent analysis of $10$~years \fermilat data \cite{Principe:2020zqe} presents the first energy-resolved morphological study at GeV energies, and suggests that the emission extends in a region larger than $2$~deg, corresponding to an intrinsic size of about $150$~pc.  
For previous analysis of this source in the radio, X-ray and TeV bands we refer to \cite{1996ApJ...466..938F,Gaensler:2002wq,Pavlov:2007av,Uchiyama:2008bd,Liu:2019vlz,Duvidovich:2019ldj,Liu:2019vlz}. All the observations in radio and X-rays provided only upper limits for the PWN emission.

\section{eHWC J1907+063 and the pulsar J1907+0602}\label{sec:J1907}
The VHE emission recently reported for this source by HAWC is significant both at energies larger than $56$~TeV ($\sqrt{TS}=10.4$) and $100$~TeV ($\sqrt{TS}=7.30$). 
The HAWC source  is centered $0.29$~deg away from the pulsar J1907+0602, and is found to be extended $0.52\pm 0.09$ deg when using a Gaussian morphology \cite{Abeysekara:2019gov}. 
The SED is better described by a log parabola with respect to a power law, and a significant emission is found up the last energy bin at  $E_\gamma>100$~TeV.

The {\it Fermi}-LAT observation of the radio-quiet $\gamma$-ray pulsar J1907.9+0602 within the TeV source extent suggested that the VHE source could be its PWN \cite{Abdo_2010}. No significant emission in the GeV range was observed in the off-peak analysis. 
The authors of Ref.~\cite{Abdo_2010} also reported a possibly extended compact X-ray source with significant non-thermal emission within the VHE extension, possibly connected to the PWN, although no other radio or X-ray measurements have confirmed its presence. 

This candidate PWN is considered to be physically more extended than other TeV PWNe of similar age (an angular extension corresponding to about $40$~pc), and the TeV spectrum does not appear to soften with distance from the pulsar, as expected from electron cooling. 
The large extent could be explained by the ICS emission produced by $e^\pm$ escaped from the nebula and diffusing in the ISM. Also, an interaction of the pulsar wind with the nearby molecular clouds in the SNR shock of SNR G40.5-0.5 has been proposed to explain the large size and the lack of spectral softening \cite{2014ApJ...787..166A}.
As suggested in Ref.~\cite{2014ApJ...787..166A}, another PWN, associated with an undetected pulsar located near the southern edge of the SNR, could contribute to the observed $\gamma$-ray emission.
Finally, in the scenario in which the VHE emission has hadronic origin instead, this source has been proposed to be among the most promising galactic neutrino emitting sources \cite{Aartsen:2018ywr}. 
In Ref.~\cite{Aartsen:2018ywr} a p-value of 0.0088 ($TS=4.5$), which does not allow to claim firmly a neutrino detection from this source and that its $\gamma$-ray emission has an hadronic origin.

The distance of this source reported in the ATNF catalog is 2.37 kpc with two different electron-density models \cite{Cordes:2002wz,2017ApJ...835...29Y}.
This distance seems thus to be very well measured and the uncertainty on it does not affect significantly our results.

\section{eHWC J2019+368 and the pulsar J2021+3651}\label{sec:sourceJ2019}
This source is located in the Cygnus region, a complex gas and star formation region in the direction of the Local Arm of our Galaxy, where tens of sources are observed at different wavelengths (see Refs.~\cite{2012A&A...538A..71A,Abeysekara_2018} and references therein). 
It is also the brightest portion of diffuse high energy $\gamma$-rays in the northern hemisphere \cite{Abdo_2007}. 
Extended emission around the pulsar J2021+3651 has been observed in X-rays (often called Dragonfly- PWN~G75.23+0.12 \cite{Van_Etten_2008}) and VHE $\gamma$-rays by different observatories, making this very-high energy source a candidate PWN. 
However, given the complicated region, contributions from hadronic processes or unrelated sources cannot be excluded, and the interpretation of the $\gamma$-ray emission around this source remains unclear. 

The pulsar J2021+3651 is among the brightest pulsars observed by {\it Fermi}-LAT. The detection of its pulsed $\gamma$-rays has been reported  in Ref.~\cite{Abdo_2009} using the first months of \fermilat data. 
An off-pulse analysis revealed no excess above the interstellar emission background, setting the putative PWN flux to be less than $10$\% of the phase-averaged emission from the pulsar \cite{Abdo_2009}.
A further search for an extended PWN around J2021+3651 using 7~years of \fermilat data \cite{Abeysekara_2018} resulted in no significant detection.  

The emission recently reported by HAWC is centered at $0.27$~deg from the pulsar J2021+3651, with $\sqrt{TS}=10.2$ (4.85) at energies $>56$~TeV ($100$~TeV), and is found to be extended $0.20\pm 0.05$ deg  when analyzing HAWC data at $E_{\rm \gamma}>56$~TeV using a Gaussian morphology \cite{Abeysekara:2019gov}. 
The spectrum extends up to $\sim100$~TeV and the SED is better described by a log parabola with respect to a power law.
The distance of this source reported in the ATNF catalog is 1.80 kpc with two different electron-density models \cite{Cordes:2002wz,2017ApJ...835...29Y}.
This distance seems thus to be very well measured and the uncertainty on it does not significantly affecting our results. 

The origin of the $\gamma$-ray emission from this source has been recently investigated in the following papers \cite{Fang:2020uiz,Liu:2020gss}.
 
\end{document}